\documentclass{article}

\usepackage{arxiv}

\usepackage[utf8]{inputenc} 
\usepackage[T1]{fontenc}    
\usepackage{hyperref}       
\usepackage{url}            
\usepackage{booktabs}       
\usepackage{amsfonts}       
\usepackage{nicefrac}       
\usepackage{microtype}      
\usepackage{lipsum}
\usepackage{graphicx}
\usepackage{float}
\usepackage{subfigure}
\usepackage{lineno}

\title{Reflectance of Silicon Photomultipliers in Linear Alkylbenzene} 

\author{
 W. Wang \\
  Nanjing University, Nanjing, 210093, China \\
   \And
 G.F. Cao\thanks{also at University of Chinese Academy of Sciences, Beijing, China. Corresponding author (caogf@ihep.ac.cn)} \\
  Institute of High Energy Physics, Beijing, 100049, China \\
     \And
 Z.Q. Xie \\
  University of Chinese Academy of Sciences, Beijing, China \\
     \And
 J. Cao \\
  Institute of High Energy Physics, Beijing, 100049, China \\
     \And
 M. Qi \\
  Nanjing University, Nanjing, 210093, China\\
  \And
 L.J. Wen \\
  Institute of High Energy Physics, Beijing, 100049, China
  }

\begin{document}
\maketitle

\begin{abstract}
Reflectance of silicon photomultipliers (SiPMs) is an important aspect to understand the large scale SiPM-based detector systems and evaluate the performance of SiPMs. We report the reflactance of two SiPMs, NUV-HD-lowCT and S14160-60-50HS manufactured by Fondazione Bruno Kessler (FBK) and Hamamatsu Photonics K.K. (HPK) respectively, in linear alkylbenzene (LAB) and in air at visible wavelengths. Our results show that the reflectance of the FBK SiPM in air varies in the range of 14\% to 23\% , depending on wavelengths and angle of incidence, which is 2 time larger than that of the HPK device. This indicates that the two manufacturers are using different designs of anti-reflective coating on SiPMs' surfaces. The reflectance is reduced by about 10\% when SiPMs are immersed in LAB, compared with that measured in air. The profiles of reflected light beams are also measured by a charge-coupled device (CCD) camera for the two SiPMs. 
\end{abstract}


\section{Introduction}
The SiPM \cite{sipm} is a novel solid-state photon sensor that has been developed to detect weak light signals and is sensitive to a single photon. The SiPM consists of multiple avalanche photodiodes (APDs) arranged in a matrix and connected in parallel. The dimensions of a single APD can vary from several micrometers to tens of micrometers, depending on the requirements of various applications. Each APD is coupled to a quenching resistor and is operated in Geiger mode by applying a reverse voltage larger than its breakdown voltage. Compared with conventional photomultiplier tubes (PMTs), the SiPM is more compact and insensitive to magnetic fields, exhibits good photon detection efficiency (PDE), and has high radio purity and low operating voltage. Due to these remarkable features, SiPMs have generated much interest with respect to a broad variety of applications, in particular, modern neutrino and dark matter projects \cite{nexo}\cite{dune}\cite{darkside}\cite{darwin}. The PDE is one of the most important aspects for evaluating SiPM performance. It can be factorized into the fill factor, light transmittance, quantum efficiency and trigger probability of avalanches. A good anti-reflective coating (ARC) design is an effective way to reduce light reflections on the SiPM surface and improve its PDE. The active area of a SiPM has a mirror-like surface and primarily produces specular reflections. However, a fraction of a diffuse component is also expected because of the microstructures on the SiPM surface, such as quenching resistors, trenches and traces used to connect the APDs. Because little information regarding the layout and material compositions of ARCs is available, it is difficult to predict the reflectance of SiPMs. For SiPMs designed to detect visible light, manufacturers usually deposit a protection layer on the SiPM surfaces for the purposes of easy handling and prevention of damage to the devices. In this case, the protection layer makes the reflections on SiPMs even more complicated. During SiPM characterization, the PDE is usually measured by using a single device, where the reflection of incident light is taken as a source of PDE loss. However, in a photodetector system, the reflected light of SiPMs might be captured by other SiPMs and contribute to the overall light collection efficiency. Therefore, understanding the reflectance of SiPMs becomes an important topic for large area SiPM-based detectors, such as the Taishan Antineutrino Observatory (TAO) experiment. This enables a better understanding of the photodetector system and SiPM devices. However, little knowledge is available regarding the reflectance of SiPMs in air or in any other working media at visible wavelengths, even though some investigations have been conducted in the ultraviolet region in vacua and in liquid xenon \cite{nexo_vac}\cite{nexo_lxe}. Commercial setups are available to measure the reflectance of samples in air at visible wavelengths; however, they cannot be used to measure the reflectance in liquids, such as LAB, the liquid used in the TAO detector. Moreover, it is difficult to predict the reflectance of SiPMs in LAB based on their reflectance results measured in air due to the aforementioned reasons of unknown ARC properties.

The TAO experiment has been proposed to precisely measure the reactor antineutrino energy spectrum. It is a satellite experiment of the Jiangmen Underground Neutrino Observatory (JUNO) \cite{juno_yb}\cite{juno_cdr}. Both JUNO and TAO are managed by the JUNO collaboration. TAO will be installed 30-35 m away from one of the reactor cores at the Taishan Nuclear Power Plant. An $\sim$2.6-ton gadolinium-doped liquid scintillator (Gd-LS) serves as the target and fills a spherical acrylic vessel with a diameter of 1.8 m. Similar to the Gd-LS used in the Daya Bay experiment \cite{dyb_ls}, LAB is chosen as the solution for the liquid scintillator in the TAO detector; however, the recipe for the Gd-LS is modified to obtain good transparency at its operating temperature (-50 degrees). An $\sim$10~$m^{2}$ SiPM array with coverage of greater than 90\% has been proposed to detect the scintillation light (350 nm to 600 nm) in the TAO detector. SiPMs with PDE greater than 50\% are required to achieve sufficient light collection efficiency and realize the designed energy resolution at the level of 1\% at 1 MeV. Because of the high dark noise rate of SiPMs at room temperature ($\sim$100 kHz/mm$^2$), the TAO detector will be cooled and operated at approximately -50 degrees so that the dark noise rate can be reduced by approximately 3 orders of magnitude \cite{sipm_dn} and its impacts on the energy resolution can be significantly suppressed. In the TAO detector, the light reflected on the SiPMs' surfaces can be easily detected again by the photodetector system due to its high coverage of photon sensors. Therefore, the equivalent PDE of SiPMs is higher than that achieved with a standalone setup measurement. Information regarding SiPM reflectance not only can lead us to a better understanding of the performance of the TAO detector but also is an important input for the detector simulation. In this work, we measure the reflectance in air and in LAB for two SiPMs. One is from FBK, model NUV-HD-lowCT, and its pixel size is 40 $\mu$m \cite{fbk_paper}. The other one is manufactured by HPK, model number S14160-6050HS, with a pixel size of 50 $\mu$m \cite{hpk_datasheet}. The measurements reported in the paper are relevant for the TAO detector. The dimensions of the two SiPMs are both $6~mm\times6~mm$. They are enclosed with epoxy resin and silicone resin as protective layers for the FBK SiPM and the HPK SiPM, respectively. 

The rest of this paper is organized as follows. First, we introduce the experimental setup, which is designed to measure the reflectance of samples in both a liquid and in air. Then, we discuss the measurement principle, validations and estimation of uncertainties. Finally, we report the reflectance results for the two SiPMs measured in air and in LAB at wavelengths from 390 nm to 600 nm and angles of incidence (AOIs) from 10 degrees to 60 degrees. 

\section{Experimental setup}
\begin{figure}[h]
    \centering
    \includegraphics[width=10cm]{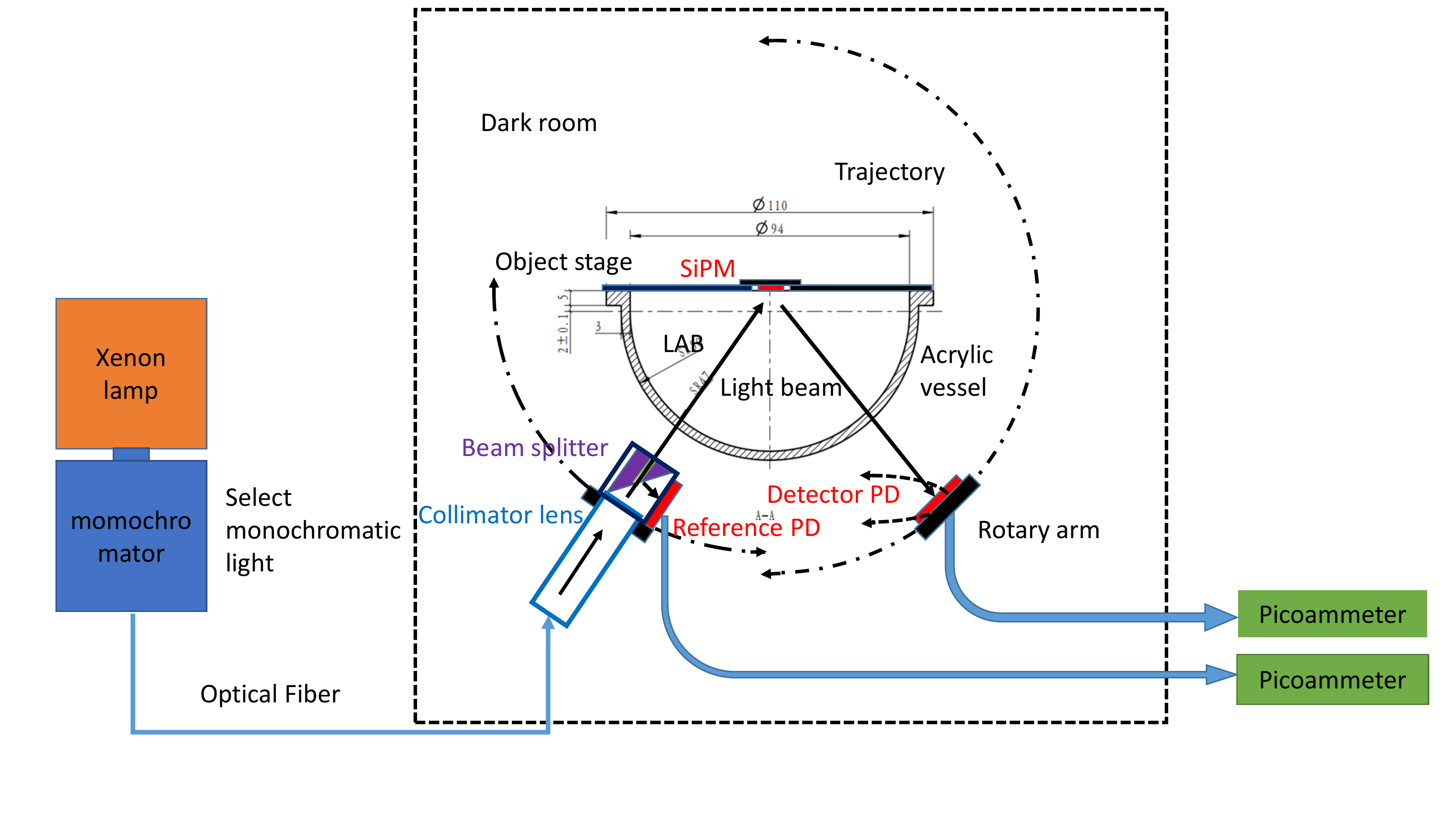}
    \caption{Schematic diagram of the instrumentation.}
    \label{setup}
\end{figure}

\begin{figure}[h]
    \centering
 \subfigure[]{
  \begin{minipage}{0.5\linewidth}
  \centering
   \includegraphics[width=6cm]{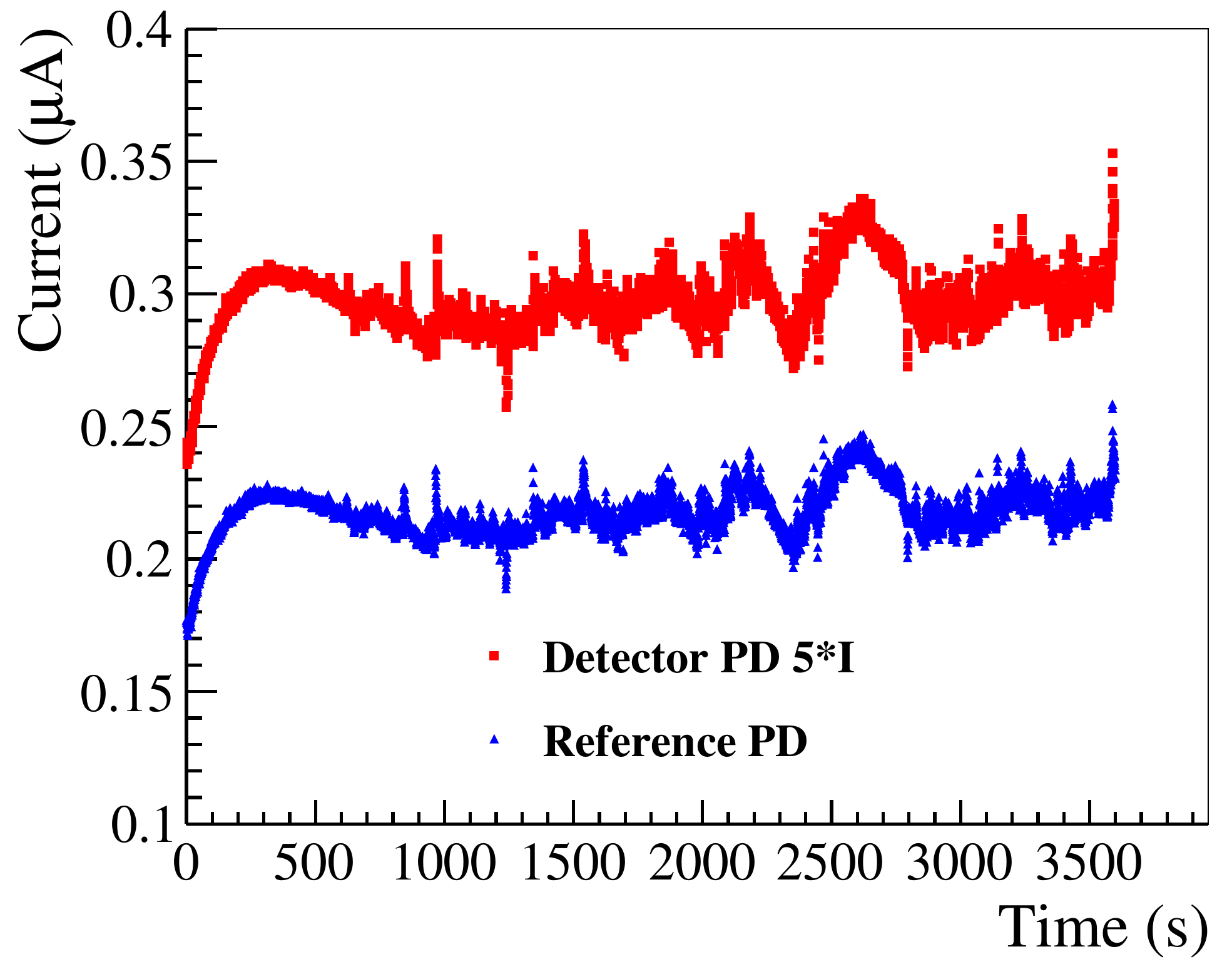}
  \end{minipage}%
  }%
   \subfigure[]{
  \begin{minipage}{0.5\linewidth}
  \centering
   \includegraphics[width=6cm]{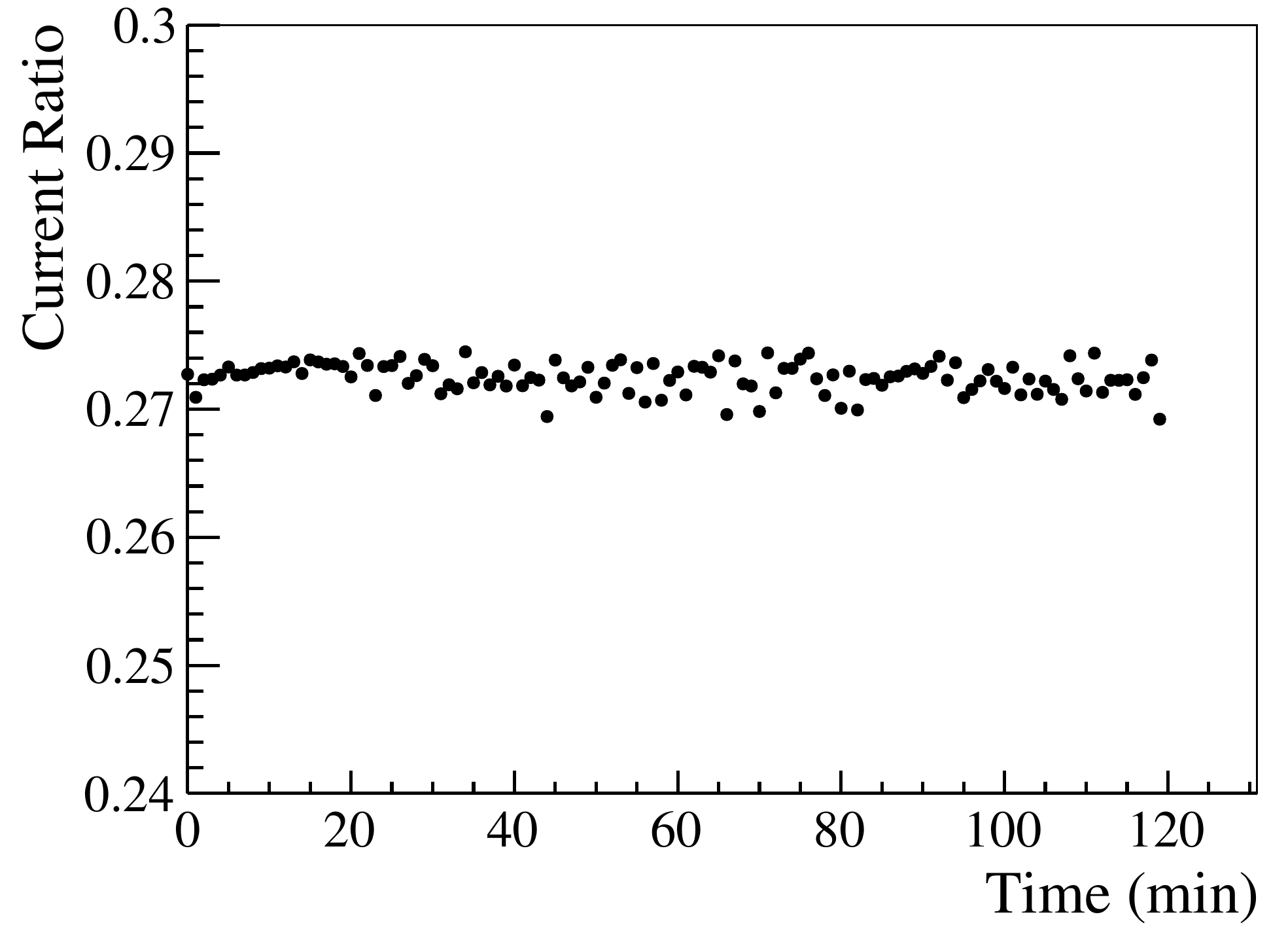}
  \end{minipage}%
  }%
    \caption{Light-intensity stability of the two light beams. (a) Current as a function of time as measured by the reference PD (blue) and the detector PD (red, scaled by a factor of 5); (b) current ratio between the reference PD and the detector PD.}
    \label{stability}
\end{figure}

The schematic diagram of the experimental setup is shown in Figure \ref{setup}. A xenon lamp serves as illumination to produce light with a continuous spectrum from $\sim$300 nm to $\sim$1200 nm. The xenon lamp module is directly coupled to a monochromator, which is used to select specific wavelengths. The widths of the entrance slit and the exit slit on the monochromator are both set to 0.5 mm, corresponding to a wavelength resolution of 2 nm. An optical fiber guides light from the monochromator into a dark room and is then connected to a collimator lens. The collimated light is divided into two light beams by a custom-made beam splitter. One light beam provides incident light for the samples to be measured. The size of this light beam is measured by a CCD camera, and its diameter is found to be $\sim$4 mm at the sample position. The other light beam directs photons onto a reference photodiode (PD; active area of 1 cm$\times$1 cm), which is attached to the beam splitter, to monitor the light-intensity stability of the xenon lamp. Both the collimator and the splitter are mounted on a rotary arm. The arm can be rotated around a reference point to achieve different AOIs. The reference point is defined as the center of the hemispherical acrylic vessel used to contain the liquid, which is indicated by the crossing point of the two dashed lines in the diagram. The dimensions of the acrylic hemisphere are labelled in the picture as well. A step-shaped lid on top of the hemisphere is used to support the samples to be measured. A hole with a diameter of 1 cm is drilled in the middle of the lid to expose the sample surface. The upper surface of the lid coincides with the horizontal dashed line shown in Figure \ref{setup}. Therefore, we can guarantee that the sample to be measured is immersed in the liquid and that the reference point can sit exactly on the sample surface. The light reflected from samples is detected by a detector PD with the same dimensions as the reference PD. The detector PD is mounted on another rotary arm. It can be rotated in two directions, as indicated by the dot-dashed lines in Figure \ref{setup}, to scan a spherical surface and measure the profile of reflected light. The distance from the reference point to the detector PD is 12.5 cm. By rotating the detector PD to the region above the acrylic hemisphere, it can directly measure the intensity of the incident light beam when the sample is moved away from the lid. The currents of the reference PD and the detector PD are measured by two picoammeters located outside the dark room. No external voltage is applied to the two PDs. A PC is used to collect data and remotely control the picoammeters. A CCD camera can also be used to measure the profile of the light reflected from samples. During reflectance measurements, the minimum AOI is $\sim$10 degree to avoid interference with the detector PD. The maximum AOI can reach $\sim$60 degrees because of the shadowing effect of the lid. However, when we conduct studies on the angular response of samples, the minimum AOI can start from 0 degrees. The light-intensity stability of the two light beams is investigated by using 430 nm light at the AOI of 0 degrees. The detector PD is rotated to the incident light beam direction. Data corresponding to the two PDs are taken simultaneously, and their currents are displayed in Figure \ref{stability} (a) as a function of time. The red line represents the current of the detector PD scaled by a factor of 5, and the blue line indicates the current of the reference PD. The currents of the two PDs, proportional to the light intensity illuminating the PDs, increase after initialization of the xenon lamp, then slightly decrease and ultimately start to fluctuate at the level of 30\% after approximately 15 minutes. However, a more important aspect is that the reference PD can effectively reproduce the fluctuation structures observed by the detector PD. This means that a much more stable incident light beam can be achieved by applying a correction based on the reference PD. The corrected light intensity is shown in Figure \ref{stability} (b), which represents the current ratio, averaged every minute, between the reference PD and the detector PD as a function of time. After the correction, the light-intensity stability of the incident light beam is better than $\pm$1.2\%. The stability at other wavelengths is also verified and is consistent with that measured at 430 nm.

\section{Measurement principle, validation and uncertainties}
\subsection{Measurement principle}
\begin{figure}[h]
  \centering
\subfigure[]{
  \begin{minipage}{0.5\linewidth}
  \centering
  \includegraphics[width=8cm]{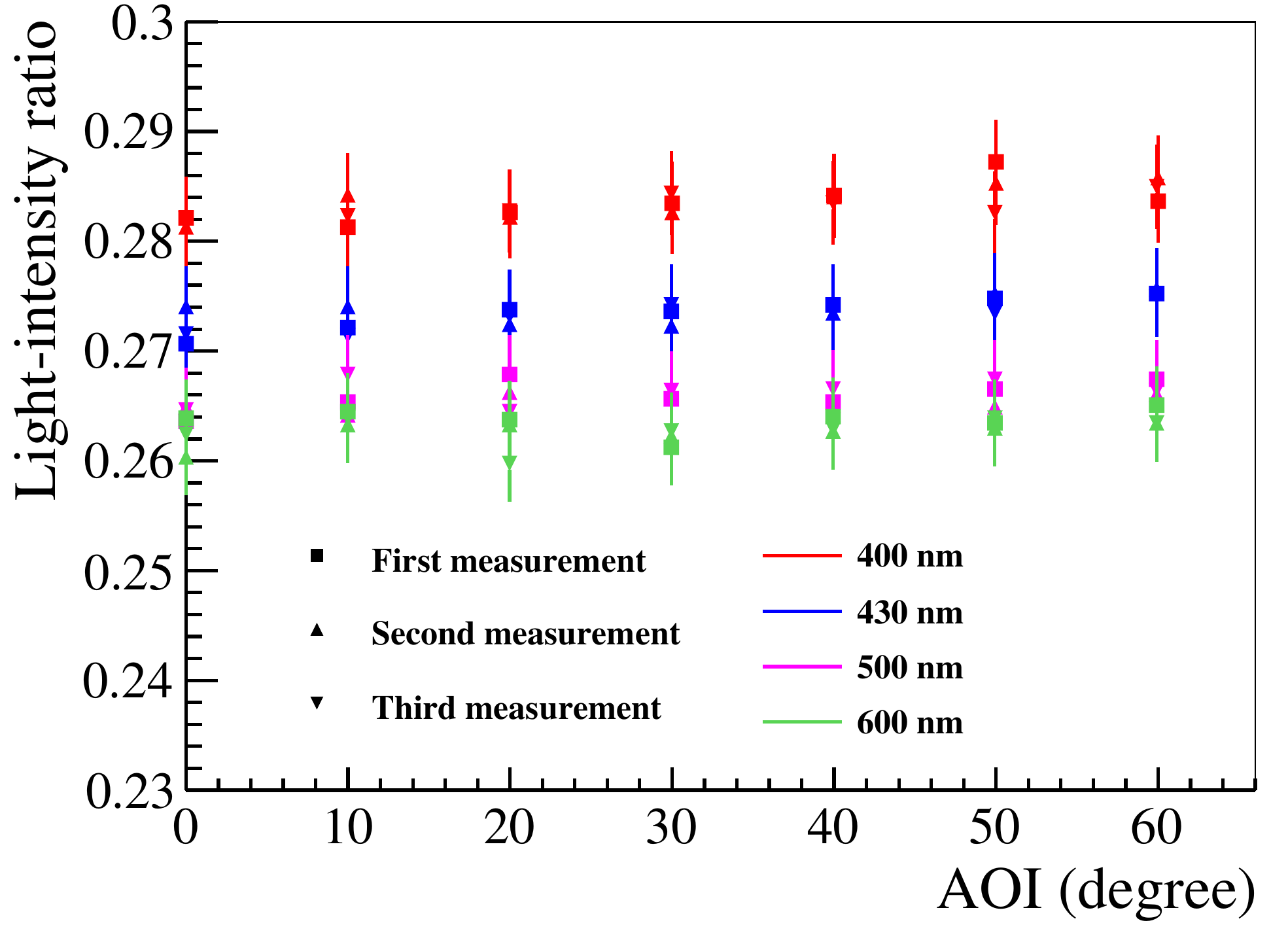}
  \end{minipage}%
  }%
  \subfigure[]{
  \begin{minipage}{0.5\linewidth}
  \centering
   \includegraphics[width=8cm]{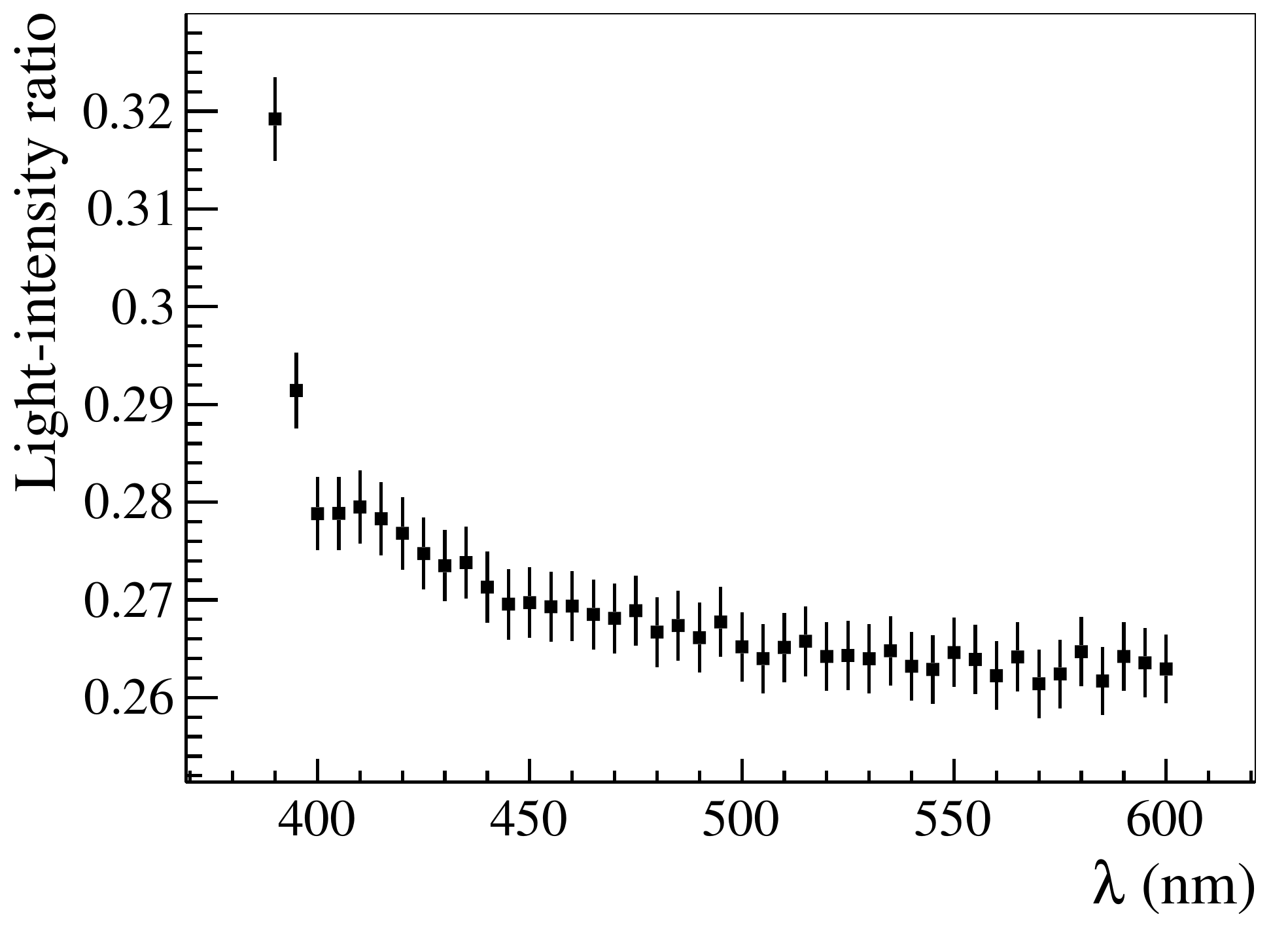}
  \end{minipage}%
  }%
 \caption{Light-intensity ratios between the reference light beam and the incident light beam. (a) Ratios as a function of AOI measured at 400 nm, 430 nm, 500 nm and 600 nm; (b) ratios as a function of wavelength measured at the AOI of 30 degrees.}
 \label{split_r_air}
\end{figure}

The reflectance of samples in air can be measured by moving the acrylic vessel away from the setup or placing an empty acrylic vessel in the setup. For the latter case, we expect a larger uncertainty in the measured reflectance since contributions from the transmittance uncertainty of the acrylic vessel and its non-uniformity must be taken into account in the reflectance measurement. Therefore, when we measure the reflectance in air for SiPMs, the acrylic vessel is moved away. However, it is deployed to measure the reflectance in air for two calibrated mirrors because it serves as a good cross-check of the results measured without the acrylic vessel. When no acrylic vessel is placed in the setup, the reflectance ($R$) can be calculated via equation \ref{eq1}.

\begin{equation}
    R = \frac{(I_d-I_d^{dark})/(I_r-I_r^{dark})}{(I_d^{0}-I_d^{dark})/(I_r^{0}-I_r^{dark})}
    \label{eq1}
\end{equation}

where $I_d$ denotes the current of the detector PD, which is measured by rotating the detector PD along the spherical surface and fixing it at the position with the maximum illumination of reflected light; $I_r$ represents the current of the reference PD. $I_d$ and $I_r$ are measured simultaneously. $I_d^{0}$ denotes the current of the detector PD, which is obtained by performing a 2D rotation for the detector PD to search for the maximum current under the illumination of the incident light; $I_r^{0}$ represents the current of the reference PD. $I_d^{0}$ and $I_r^{0}$ are also measured simultaneously. $I_d^{dark}$ and $I_r^{dark}$ indicate the dark current of the detector PD and the reference PD, respectively, which are measured by turning off the xenon lamp at the end of data collection. The light-intensity ratios between the reference light beam and the incident light beam are carefully investigated at 4 selected wavelengths (400 nm, 430 nm, 500 nm and 600 nm) and different AOIs. The results are shown in Figure \ref{split_r_air} (a). 
For each wavelength, as indicated by different colors, the measurements are repeated 3 times, shown by different markers. No significant changes are observed for the light-intensity ratios during rotation of the collimator and the splitter. The maximum relative variation is $\pm$1.34\%. To be conservative, in the following studies, we consider this number to be the uncertainty in the light-intensity ratios and assume that the ratios do not depend on the AOI. A detailed wavelength scan is performed with a step length of 5 nm in order to obtain the light-intensity ratios as a function of wavelength, as shown in Figure \ref{split_r_air} (b). In this measurement, the AOI is fixed at 30 degrees. The uncertainty for each data point is assigned according to the aforementioned $\pm$1.34\%, indicated by the size of the error bars. 

\begin{figure}[h]
  \centering
  \subfigure[]{
  \begin{minipage}{0.5\linewidth}
  \centering
  \includegraphics[width=8cm]{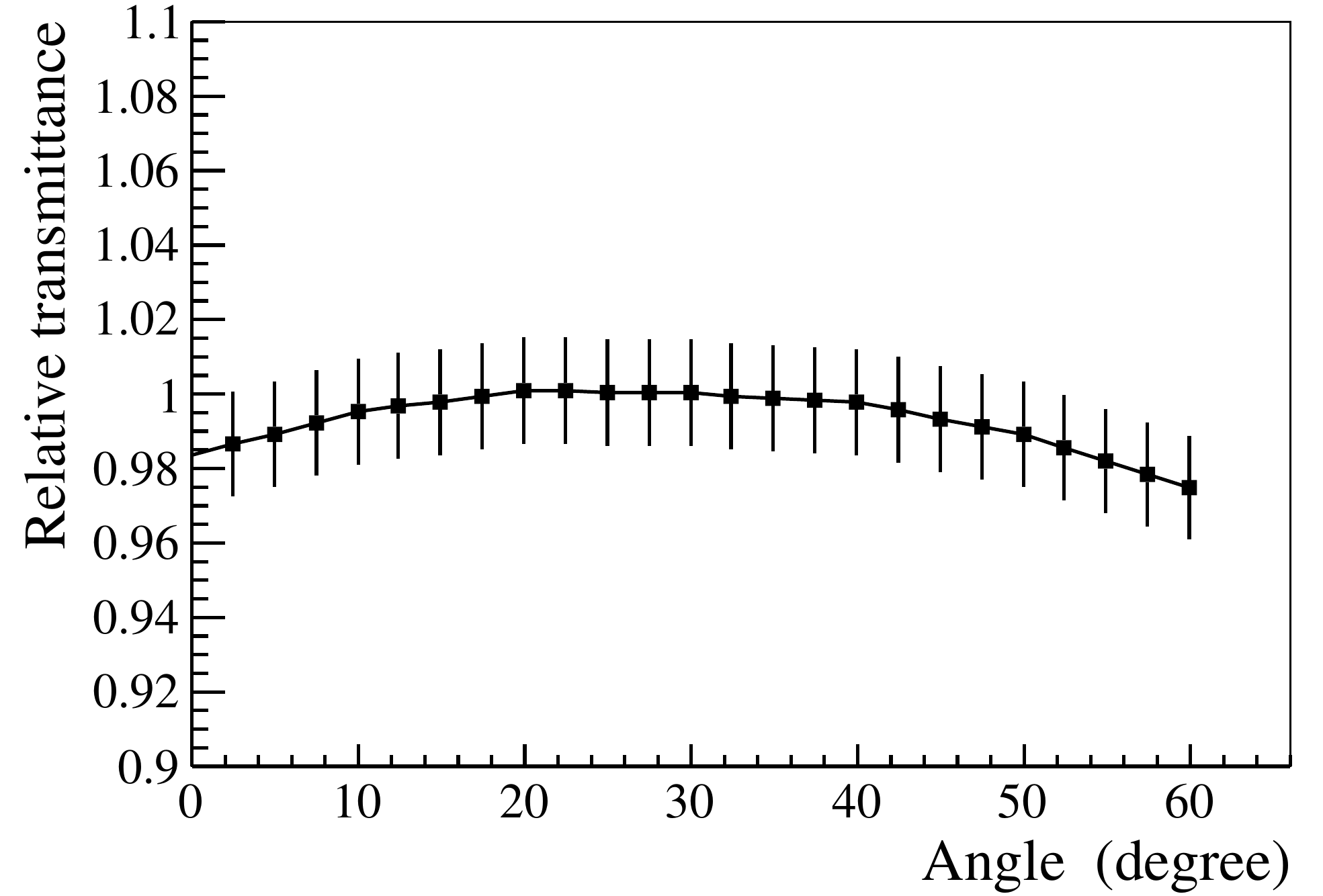}
  \end{minipage}%
  }%
  \subfigure[]{
  \begin{minipage}{0.5\linewidth}
  \centering
   \includegraphics[width=8cm]{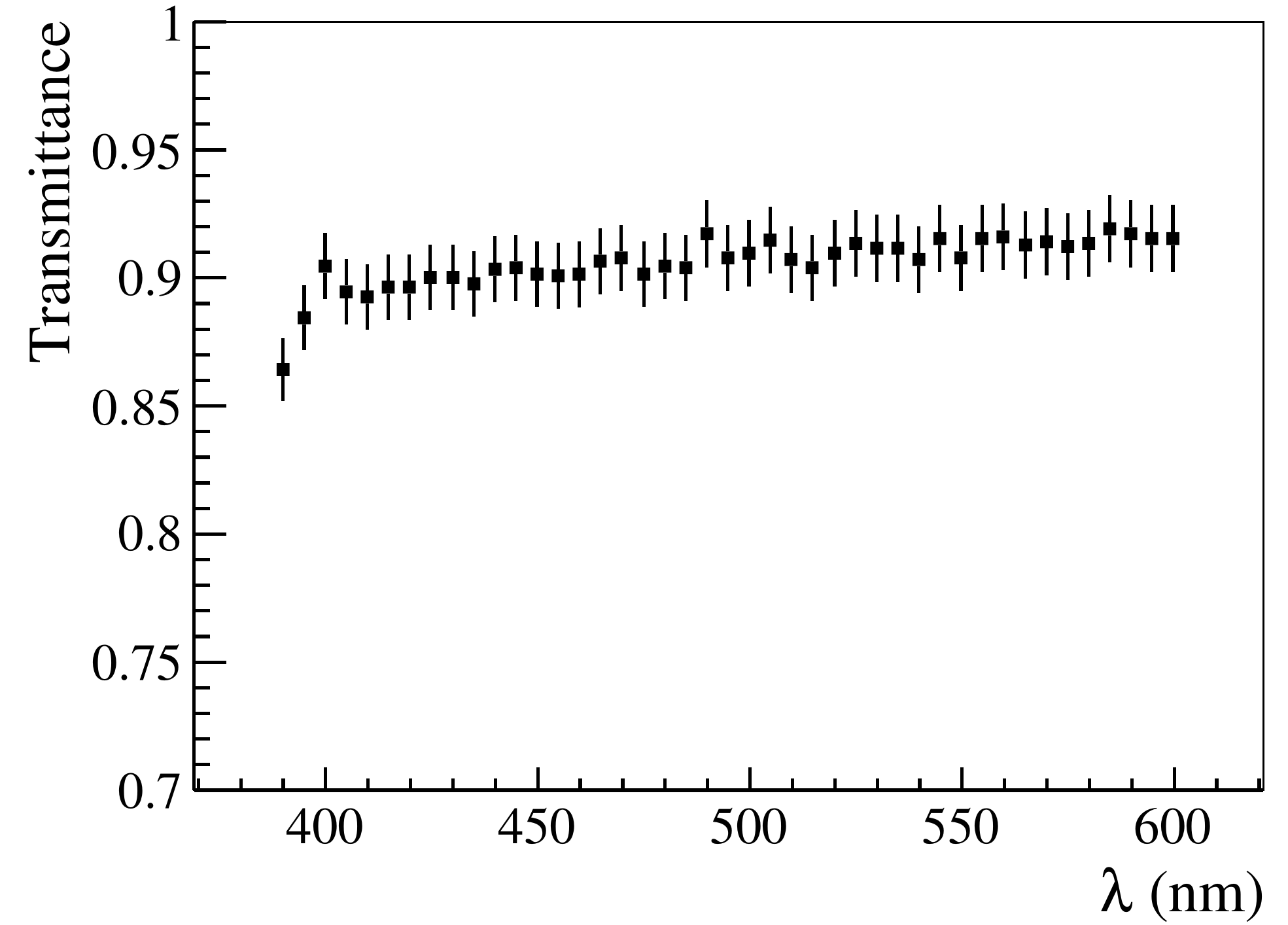}
  \end{minipage}%
  }%
 \caption{(a) Relative transmittance of the acrylic vessel as a function of AOI, where each data point is averaged with respect to 4 wavelengths and 3 randomly selected positions on the vessel; (b) transmittance of the acrylic vessel as a function of wavelength, measured at the AOI of 30 degrees.}
 \label{tran_acrylic}
\end{figure}

For the case with an empty acrylic vessel, the transmittance of the acrylic vessel must be corrected, and then, $R$ can be extracted with the new equation \ref{eq2}.

\begin{equation}
    R = \frac{(I_d-I_d^{dark})/(I_r-I_r^{dark})}{T^{2}(I_d^{0}-I_d^{dark})/(I_r^{0}-I_r^{dark})}
    \label{eq2}
\end{equation}

where $T$ denotes transmittance of the acrylic vessel in air, including reflections on surfaces of the vessel. With the input of the light-intensity ratio in the absence of the acrylic vessel, shown in Figure \ref{split_r_air} (b), $T$ is calculated by measuring the light-intensity ratio again, but with the deployment of the acrylic vessel. At each AOI, the transmittance is measured at 4 selected wavelengths (390 nm, 400 nm, 430 nm and 500 nm) and at 3 randomly selected positions on the vessel. To obtain the AOI response of the transmittance, we assume all wavelengths follow the same shape of AOI response of the transmittance, and then we normalize the absolute value of the transmittance at the AOI of 30 degrees to 1 for different wavelengths. Finally, the mean transmittance is extracted by averaging all data points at each AOI. Figure \ref{tran_acrylic} (a) shows the relative transmittance of the acrylic vessel as a function of the AOI. Its uncertainty is estimated by using the maximum variation among the aforementioned measurements, which is found to be primarily contributed by the instability of the setup and the non-uniformity of the acrylic transmittance. The angular dependence of the acrylic transmittance is caused by the technologies utilized to produce the acrylic vessel. In this work, we employ thermal bending technology, which can be improved by using casting technology to eliminate the angular dependence of the transmittance. The spectral dependence of the transmittance is measured at the AOI of 30 degrees, and the results are displayed in Figure \ref{tran_acrylic} (b), with a relative uncertainty of $\pm$1.43\%. 

\begin{figure}[h]
  \centering
   \includegraphics[width=10cm]{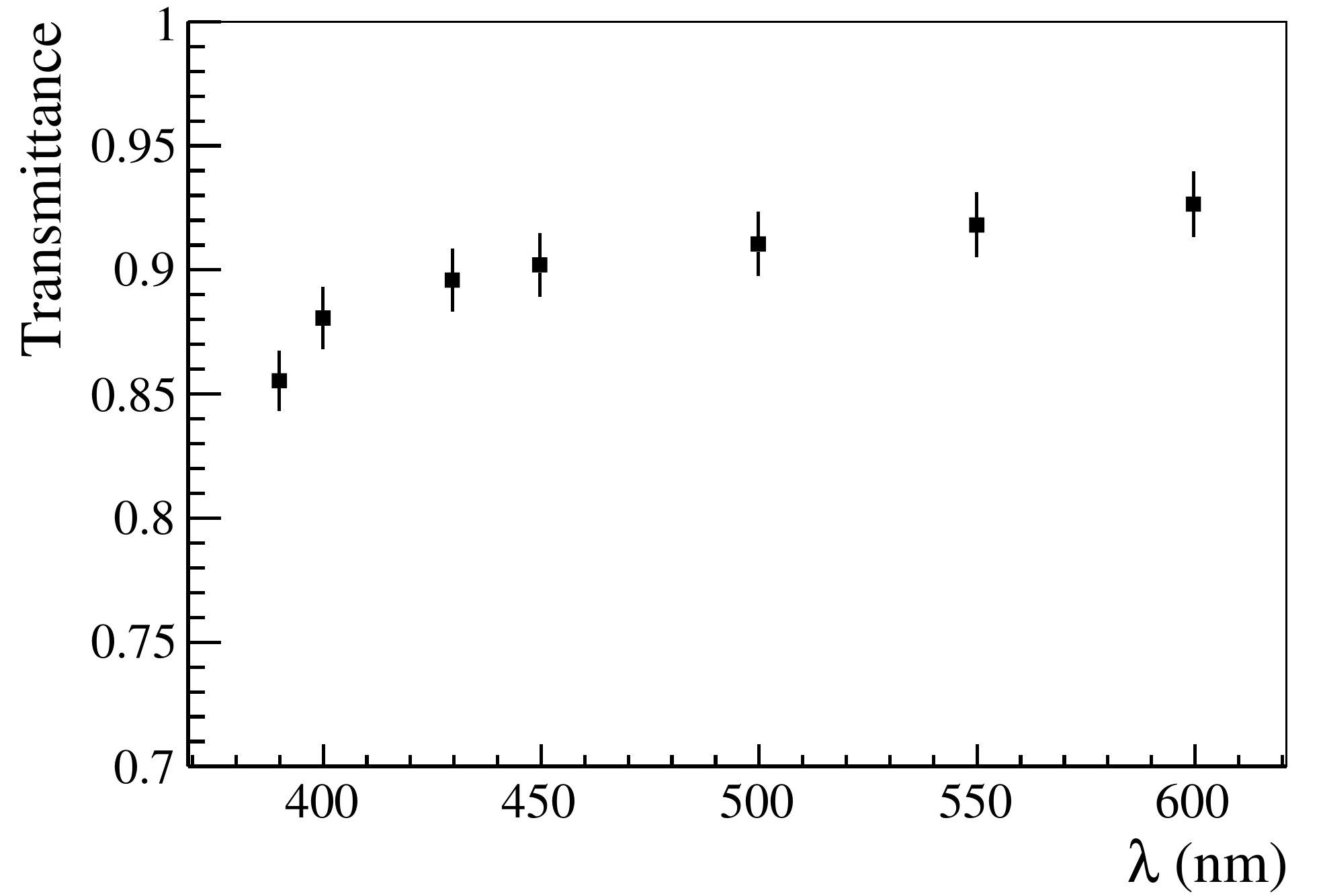}
 \caption{Transmittance of the acrylic vessel filled with LAB as a function of wavelength measured at AOI of 0 degrees, in which reflections at the LAB surface have been corrected.}
 \label{corrected_trans}
\end{figure}

To measure the reflectance of samples in a liquid, equation \ref{eq2} is still suitable for calculating $R$; however, $T$ should be replaced by $T^\prime$, which includes additional contributions from the absorption and scattering of the liquid. However, for a filled acrylic vessel, since the refraction occurs at the interface between the liquid and air and it is difficult to detect refracted light with the detector PD for non-zero AOI, the transmittance can only be measured with the setup at the AOI of 0 degrees. The AOI dependence of $T^\prime$ is determined by the acrylic vessel and follows the same trend as that of $T$ since the liquid has a much better uniformity than the acrylic vessel, and the length of the optical path in the liquid does not rely on the AOI. Similar to the $T$ measurement, when the acrylic vessel is filled by LAB, the transmittance is also measured by the reference PD and the detector PD. Reflections at the LAB surface are corrected in the measured transmittance since samples are immersed in LAB. The reflections are calculated at the AOI of 0 degrees for wavelengths of interest based on Fresnel's equations \cite{fresnel} and the refractive indices of LAB and air \cite{dyb_ls_r}. The results of $T^\prime$ versus wavelength are shown in Figure \ref{corrected_trans}, after the corrections for reflections at the LAB surface. Reflections at the interface between LAB and the acrylic are negligible since they have similar refractive indices. We assign the same uncertainty to $T^\prime$ as that of $T$ since the information of the LAB refractive indices is known with better precision and the calculated correction factors for $T^\prime$ yield negligible uncertainties. Combined with the results in Figure \ref{corrected_trans} and Figure \ref{tran_acrylic} (a), $T^\prime$ can be interpolated at any given wavelength and AOI. It can then be directly used in equation \ref{eq2} to evaluate the reflectance of samples in LAB.

\subsection{Validation}
\begin{figure}[h]
  \centering
  \subfigure[]{
  \begin{minipage}{0.5\linewidth}
  \centering
   \includegraphics[width=7cm]{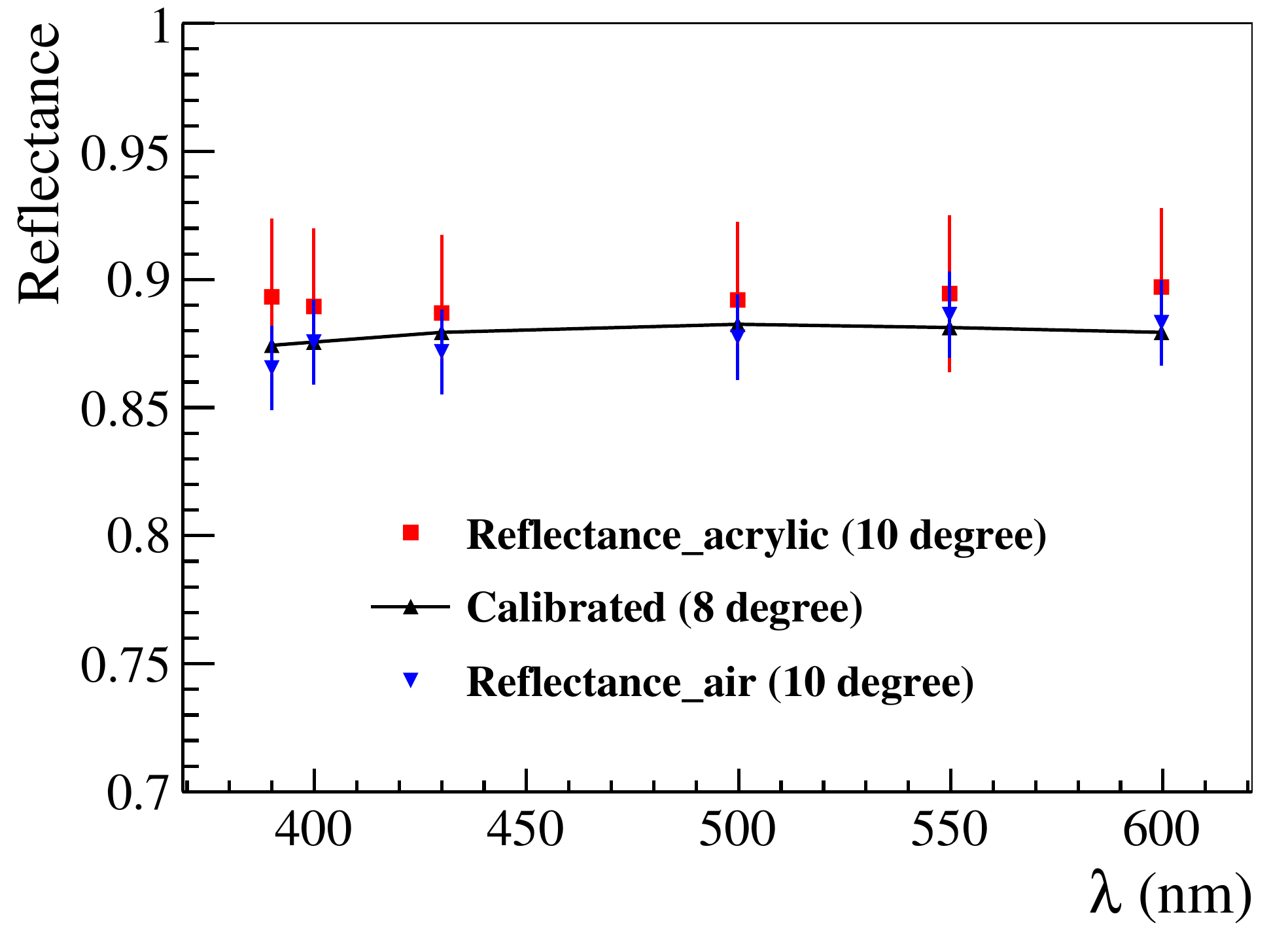}
  \end{minipage}%
  }%
  \subfigure[]{
  \begin{minipage}{0.5\linewidth}
  \centering
   \includegraphics[width=7cm]{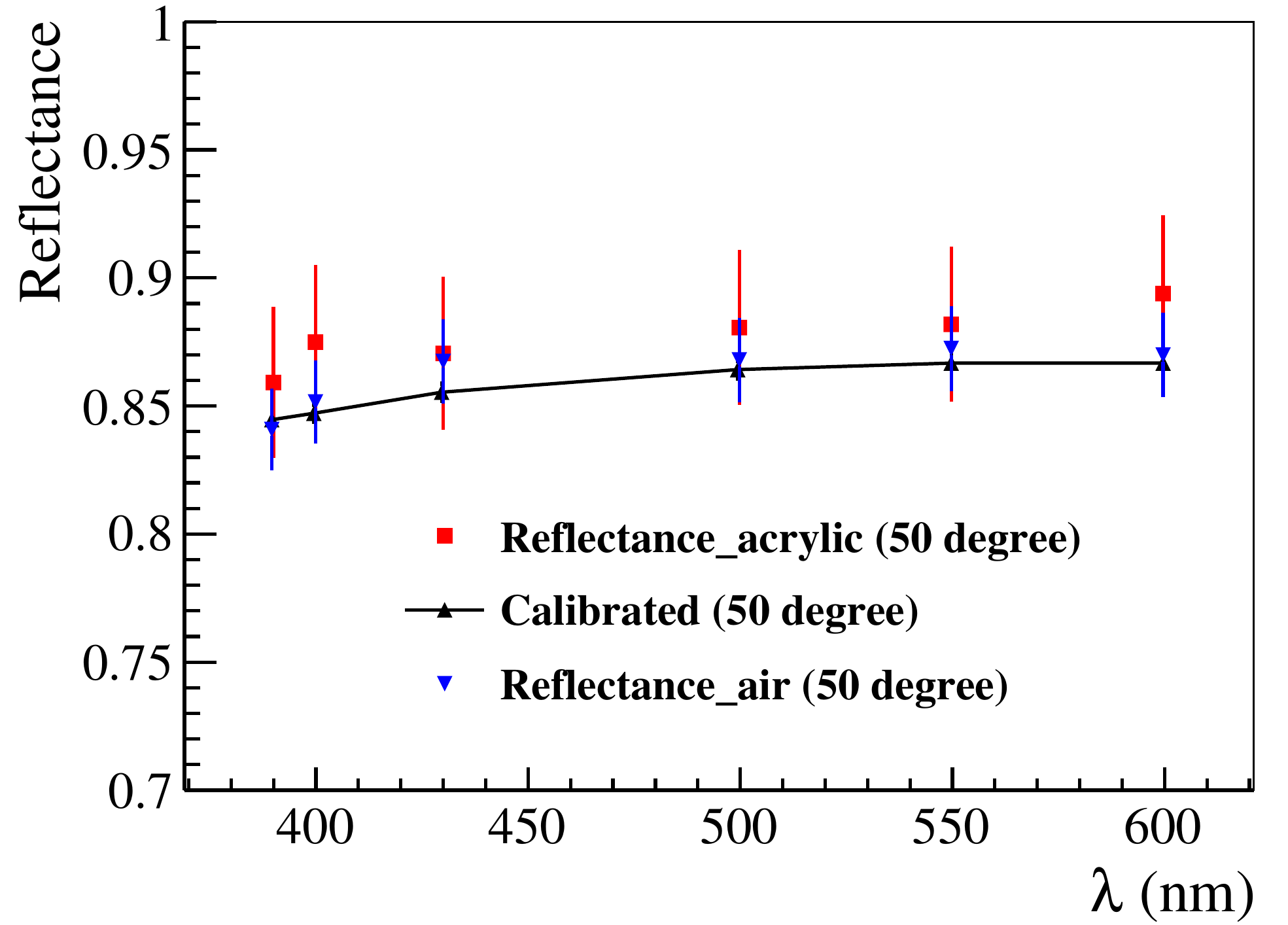}
  \end{minipage}%
  }%
 \caption{Reflectance of two reference mirrors as a function of wavelength measured using our setup with (red markers) and without (blue markers) the acrylic vessel at AOIs of 10 degrees (a) and 50 degrees (b), compared with calibrated results (black lines) at AOIs of 8 degrees (a) and 50 degrees (b).}
 \label{ref_mirror}
\end{figure}

We use two metal-coated plane mirrors, which are available in our lab, to serve as two reference mirrors and validate the results obtained by our experimental setup. The reflectance values of the reference mirrors in air are calibrated by the National Institute of Metrology in China for a spectral range from 350 nm to 900 nm. The AOI is fixed at 8 degrees and 50 degrees for the two mirrors. The relative uncertainty of the calibration is 0.5\%. By using our setup, the reflectance values of the two mirrors in air are measured again at six selected wavelengths. Since our setup cannot reach the AOI of 8 degrees, the AOI of 10 degrees is used instead for the reference mirror calibrated at 8 degrees. The angular response of the reflectance is investigated via Fresnel's equations and the optical properties of the metal film stack on the mirror. The results show that the reflectance difference between the two AOIs is negligible compared to the uncertainties induced by the setup. Figure \ref{ref_mirror} presents the measured reflectance in air (shown by blue markers) as a function of wavelength for AOIs of 10 degrees (a) and 50 degrees (b), compared with calibration curves (represented as black lines). The reflectance measured with an empty acrylic vessel (indicated by red markers) is also shown in the plot for verification of the consistency. The results measured by our setup agree well with the calibrated results within the measured uncertainties for the two reference mirrors. The reflectance measured with the empty acrylic vessel is consistent with that measured without the vessel; however, its uncertainty becomes larger due to the aforementioned reasons related to the transmittance non-uniformity of the acrylic vessel. It is difficult to perform validations for reflectance measured in liquids due to the lack of reference samples calibrated in liquids.

\subsection{Estimation of uncertainties}
In equation \ref{eq1}, the uncertainty in $R$ is estimated by performing an error propagation with inputs of the uncertainties in the light-intensity ratios in the numerator and denominator. They are both determined by the maximum fluctuation in Figure \ref{split_r_air} (a), which is found to be $\pm$1.34\%. This yields a relative uncertainty in $R$ of 1.9\%. To be conservative, we assume that no correlations exist between the light-intensity ratios in the numerator and denominator. The instability of the light beam emitted from the xenon lamp is one of the main factors contributing to the uncertainty in $R$. Another major factor is the fluctuation of the current ratio of the two PDs during rotation of the incident light beam. This is caused by the imperfect splitter and non-uniformly distributed light intensities in the cross section of the light beams. Other factors, such as the accuracy of the picoammeters and temperature stability of PDs, are negligible. For $R$ in liquid, the transmittance in equation \ref{eq2} must be considered. Its uncertainty is estimated to be $\pm$1.43\%, as discussed in section 3.1. Using the same aforementioned uncertainties for other parameters, the relative uncertainty in $R$ in a liquid is calculated as 3.4\%.

\section{Results}
\subsection{Specular reflectance}
\begin{figure}[h]
  \centering
\subfigure[FBK SiPM in air]{
  \begin{minipage}{0.5\linewidth}
  \centering
  \includegraphics[width=8cm]{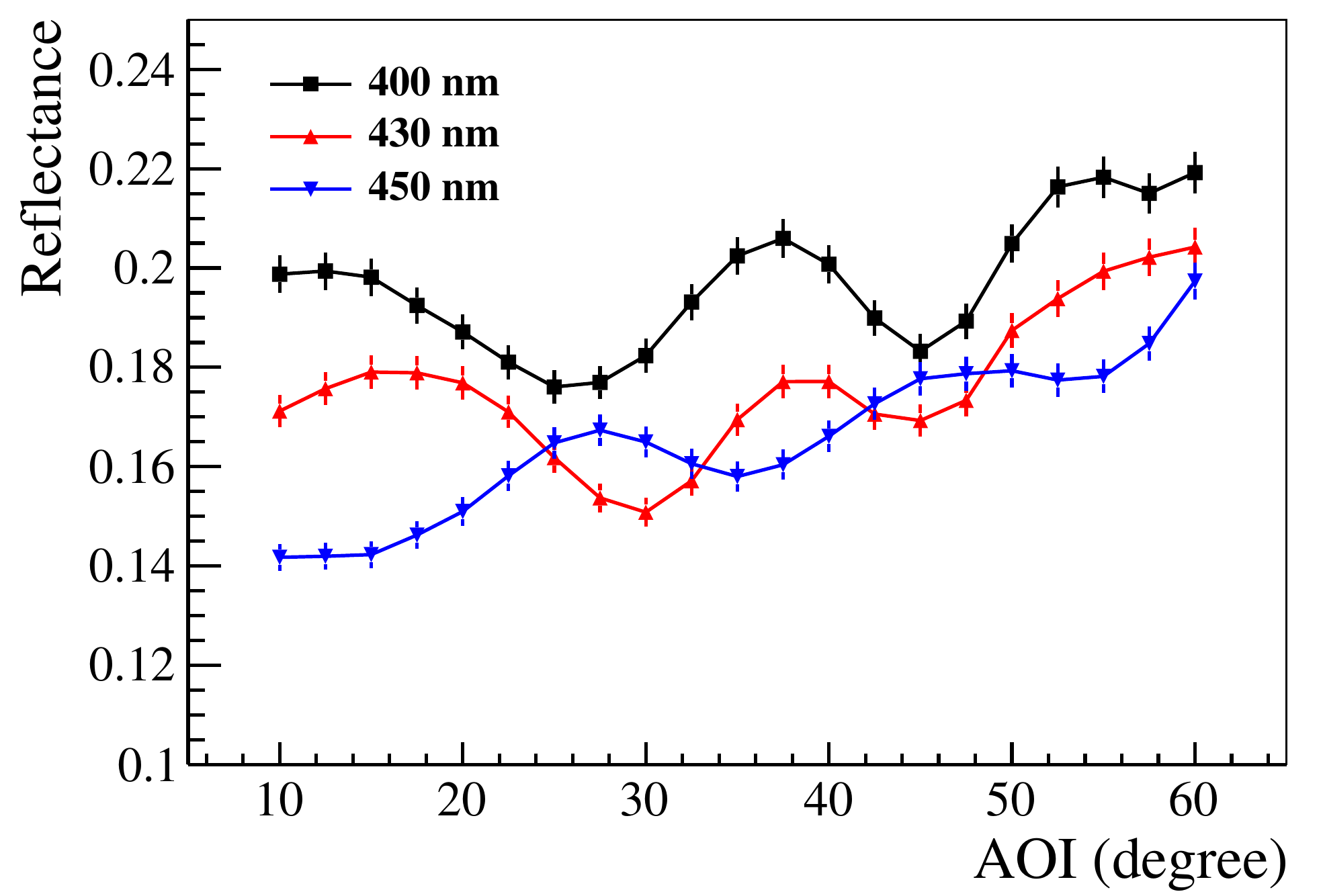}
  \end{minipage}%
  }%
  \subfigure[FBK SiPM in LAB]{
  \begin{minipage}{0.5\linewidth}
  \centering
   \includegraphics[width=8cm]{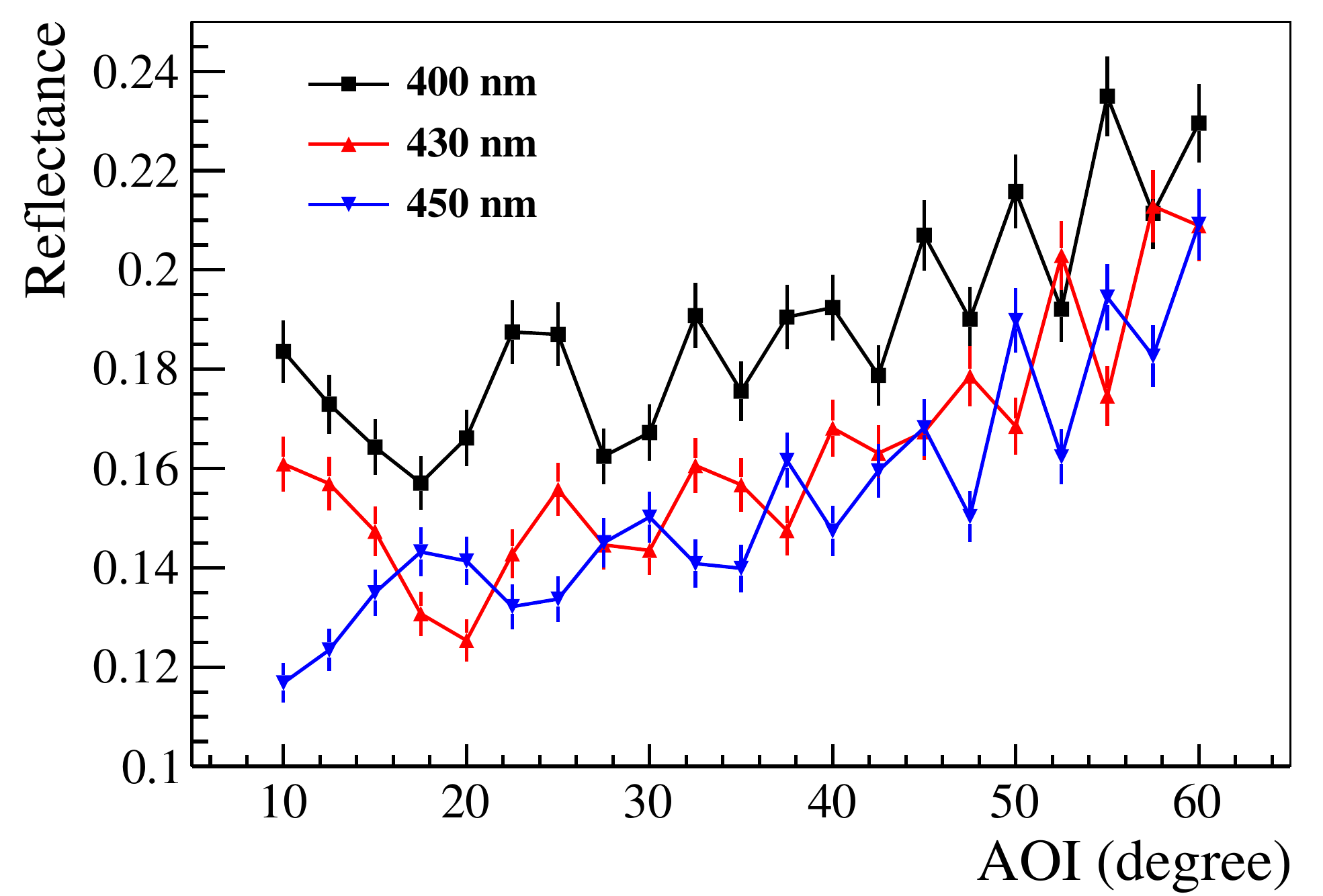}
  \end{minipage}%
  }%
  
  \subfigure[HPK SiPM in air]{
  \begin{minipage}{0.5\linewidth}
  \centering
  \includegraphics[width=8cm]{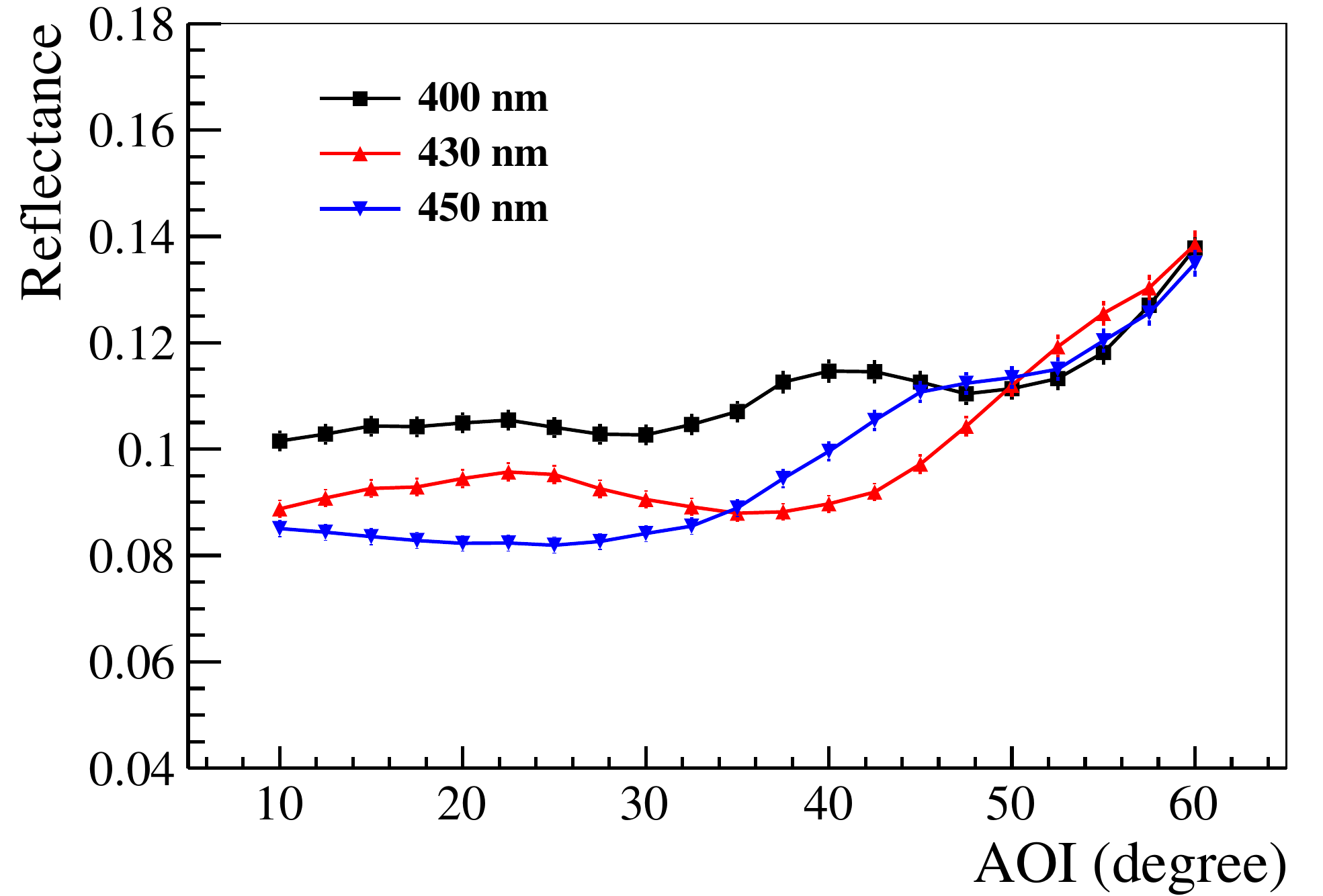}
  \end{minipage}%
  }%
  \subfigure[HPK SiPM in LAB]{
  \begin{minipage}{0.5\linewidth}
  \centering
   \includegraphics[width=8cm]{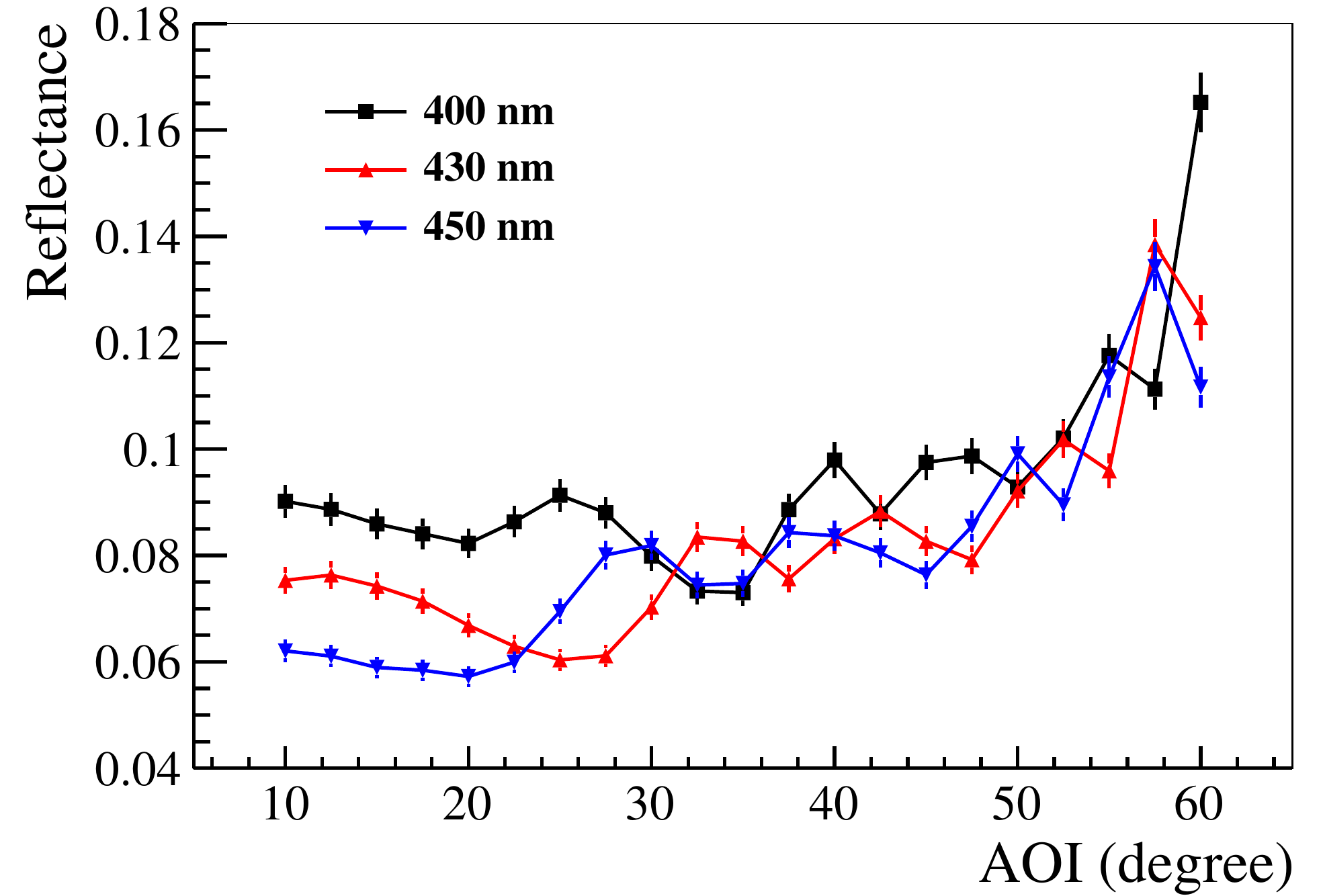}
  \end{minipage}%
  }%
 \caption{Specular reflectance in air and LAB of the FBK SiPM and HPK SiPM as a function of AOI, measured at wavelengths of 400 nm, 430 nm and 450 nm.}
 \label{spec_angle}
\end{figure}

The angle dependence of the specular reflectance of the two SiPMs is measured at wavelengths of 400 nm, 430 nm and 450 nm. The selected wavelengths are in the spectral region of the liquid scintillator with bright fluorescence. The AOI is scanned from 10 degrees to 60 degrees with a step of 2.5 degrees for each wavelength and each SiPM. The results of the reflectance in air and in LAB are shown in Figure \ref{spec_angle} as a function of AOI. In general, the two SiPMs are more reflective in air than in LAB at the level of approximately 10\%. The excess reflectance is partially contributed by the protection layer of the SiPMs. When SiPMs are placed in air, approximately 5\% of the light is reflected by the surface of the protection layer at the AOI of 10 degrees, and this can increase to 15\% at the AOI of 60 degrees. This calculation is based on the refractive index of the protection layer, without considering its wavelength dependence. Its value is 1.57, as provided in HPK's datasheet \cite{hpk_datasheet}. However, the reflections occurring on the protection layer surface are significantly suppressed and become negligible when the SiPMs are immersed in LAB. The measured reflectance is a result of the combination of the protection layer and ARC on the SiPMs. However, the contribution of the ARC is difficult to predict due to the lack of ARC information. The reflectance of the FBK SiPM is almost 2 times that of the HPK SiPM, which is mainly caused by the different ARC designs and layouts used in the FBK and HPK SiPMs, rather than to the protection layers because similar optical properties are expected for epoxy resin (used in the FBK SiPM protection layer) and silicone resin (used for the HPK SiPM). Oscillation structures are observed both in air and in LAB for the two SiPMs, which are by the interference of reflected light in the ARC stacks on the surfaces of the two SiPMs. Thay are used to improve the transmittance. Our results also show that oscillations are suppressed when we operate the SiPMs in LAB. 

\begin{figure}[h]
  \centering
\subfigure[FBK SiPM in air]{
  \begin{minipage}{0.5\linewidth}
  \centering
  \includegraphics[width=8cm]{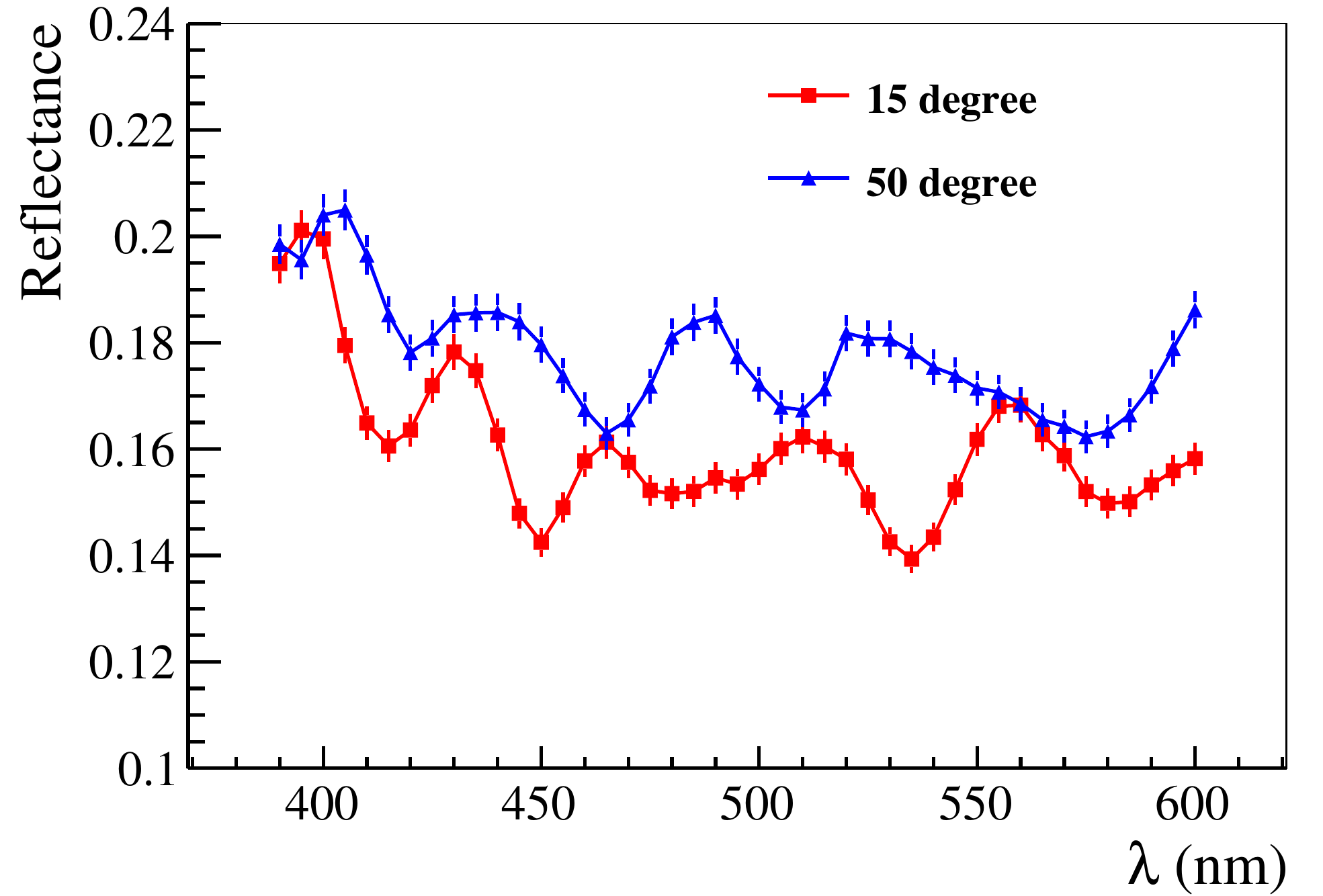}
  \end{minipage}%
  }%
  \subfigure[FBK SiPM in LAB]{
  \begin{minipage}{0.5\linewidth}
  \centering
   \includegraphics[width=8cm]{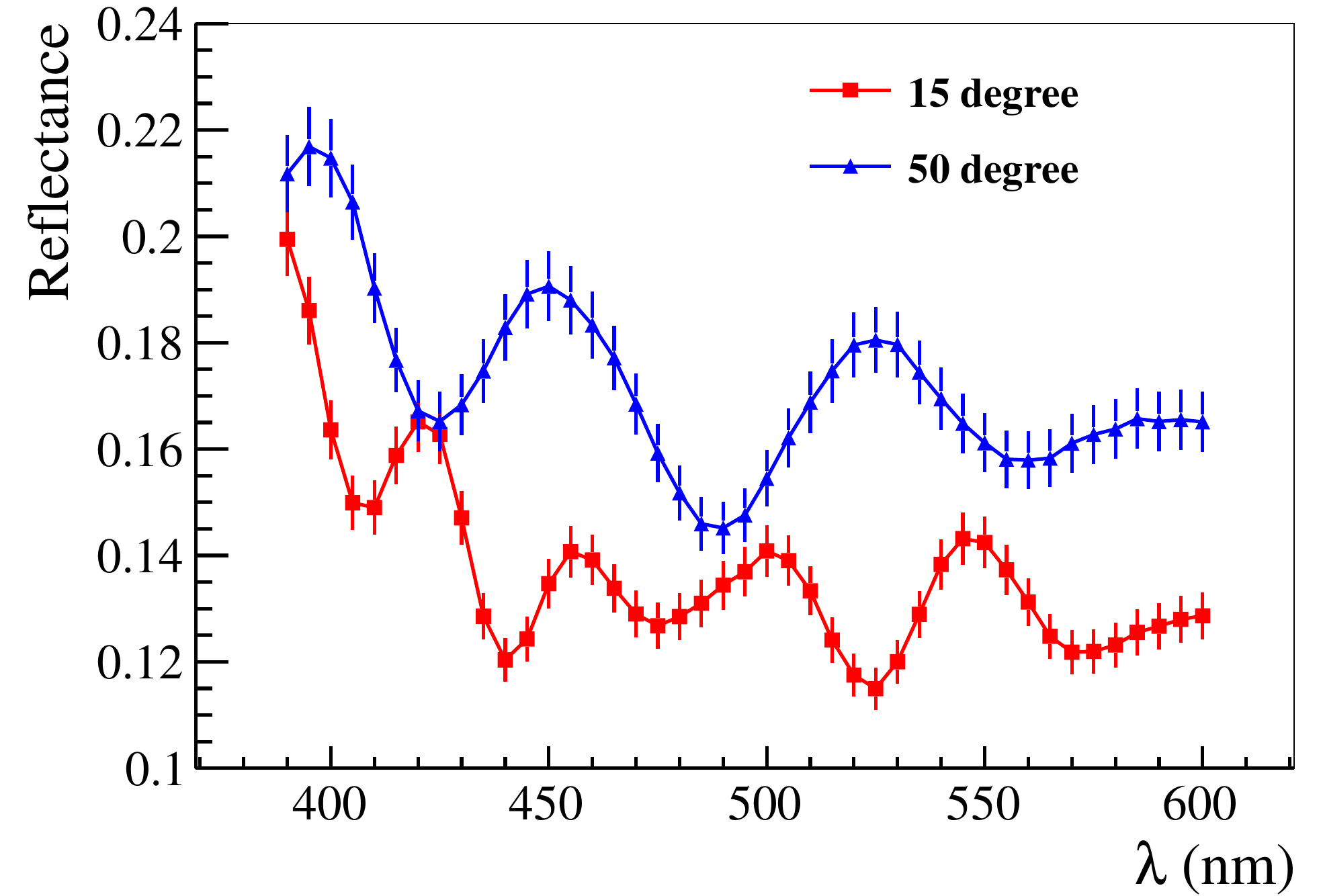}
  \end{minipage}%
  }%
  
  \subfigure[HPK SiPM in air]{
  \begin{minipage}{0.5\linewidth}
  \centering
  \includegraphics[width=8cm]{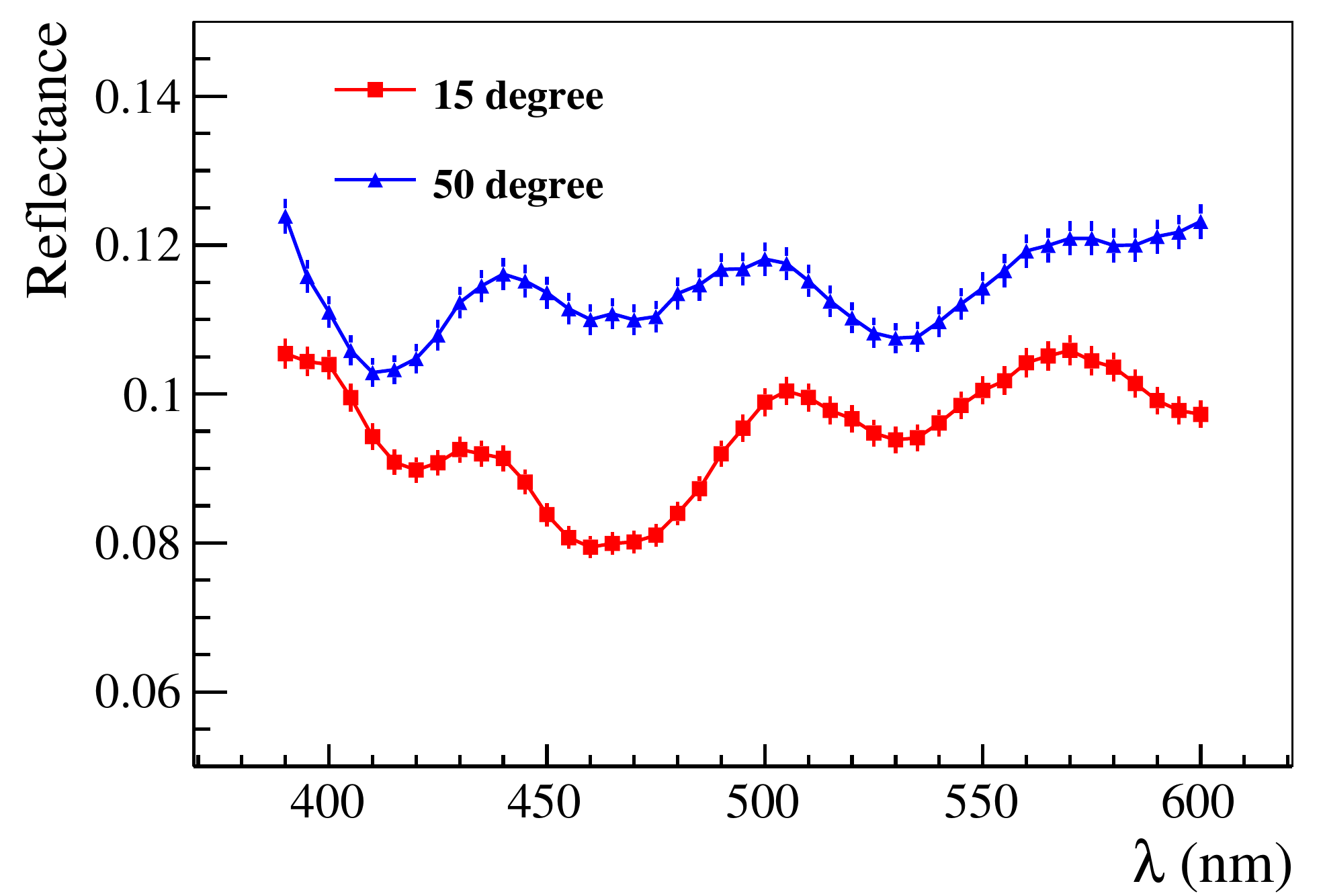}
  \end{minipage}%
  }%
  \subfigure[HPK SiPM in LAB]{
  \begin{minipage}{0.5\linewidth}
  \centering
   \includegraphics[width=8cm]{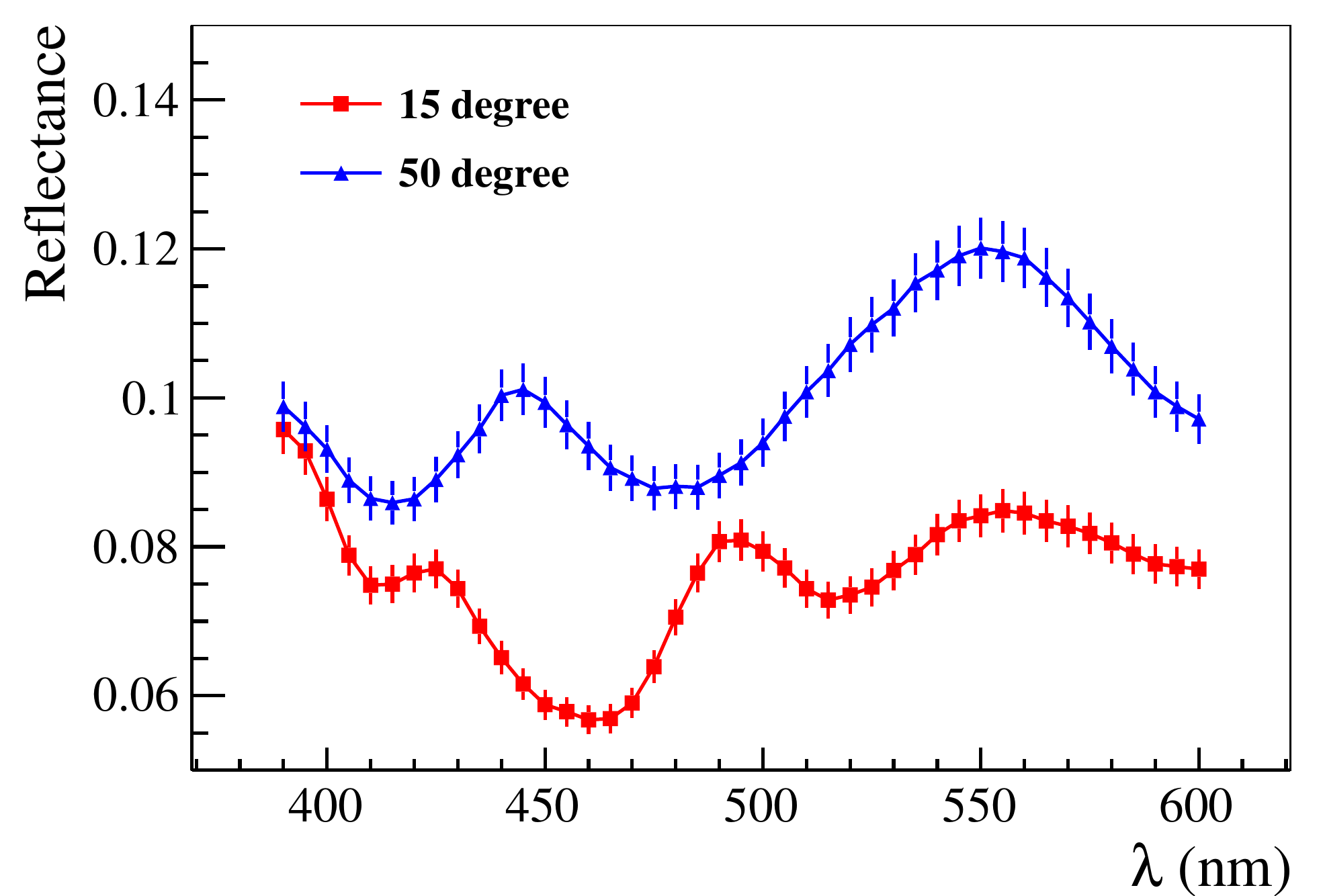}
  \end{minipage}%
  }%
 \caption{Specular reflectance in air and LAB of the FBK SiPM and the HPK SiPM as a function of wavelength, measured at AOIs of 15 degrees and 50 degrees.}
 \label{spec_wl}
\end{figure}

Figure \ref{spec_wl} represents the specular reflectance of the two SiPMs as a function of wavelength, measured in air and in LAB at AOIs of 15 degrees (red) and 50 degrees (blue). The measured wavelengths cover most of the spectral region of scintillation light from the liquid scintillator, except for short wavelengths from 350 nm to 390 nm because the light intensities at short wavelengths provided by the setup are not sufficient to guarantee an accurate reflectance measurement. The relative reflectance uncertainties in air and in LAB are assigned as 1.9\% and 3.4\%, respectively, as discussed in section 3.3. The FBK SiPM is more reflective than the HPK SiPM in the entire measured spectral range. The reflectance values of the two SiPMs also oscillate according to the wavelength, which is caused by the same reasons as those shown in Figure \ref{spec_angle}. In LAB, the oscillation cycle becomes longer at the AOI of 50 degrees compared with that in air. However, at the AOI of 15 degrees, the cycles are nearly identical according to the results in LAB and in air, and they share similar oscillation structures. 

\subsection{Diffuse reflectance}
\begin{figure}[h]
  \centering
\subfigure[FBK SiPM in air]{
  \begin{minipage}{0.5\linewidth}
  \centering
  \includegraphics[width=8cm]{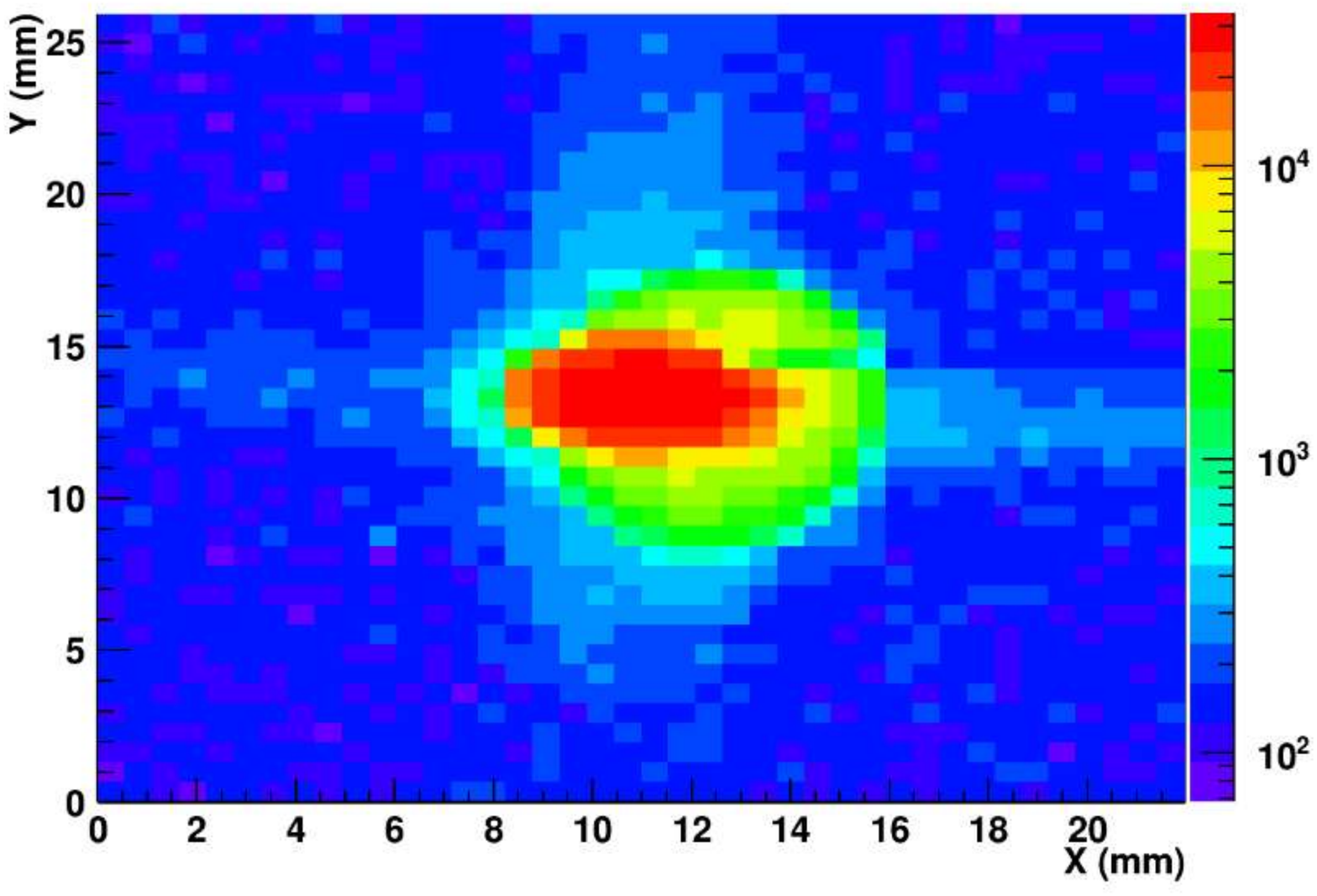}
  \end{minipage}%
  }%
  \subfigure[FBK SiPM in LAB]{
  \begin{minipage}{0.5\linewidth}
  \centering
   \includegraphics[width=8cm]{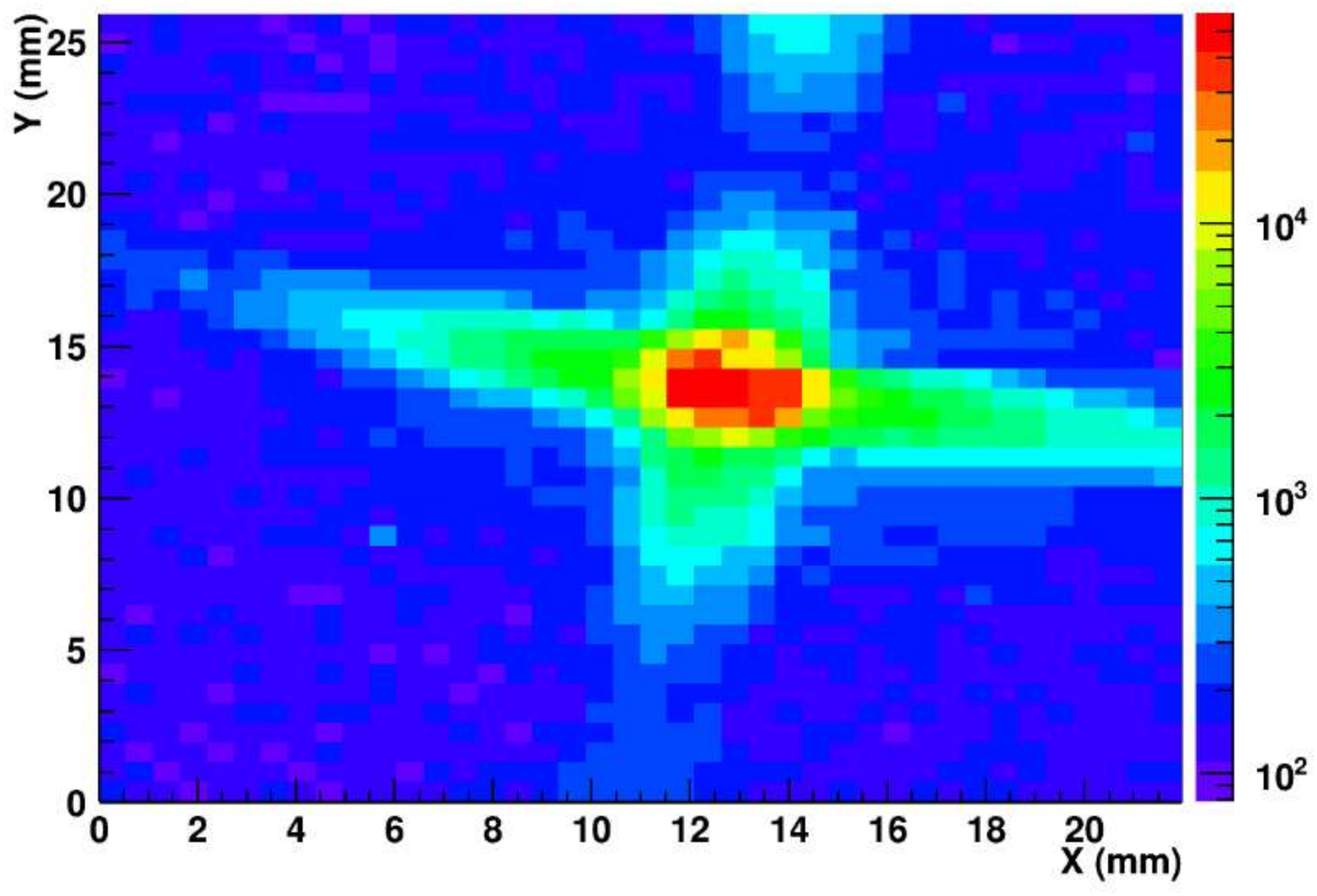}
  \end{minipage}%
  }%
  
  \subfigure[HPK SiPM in air]{
  \begin{minipage}{0.5\linewidth}
  \centering
  \includegraphics[width=8cm]{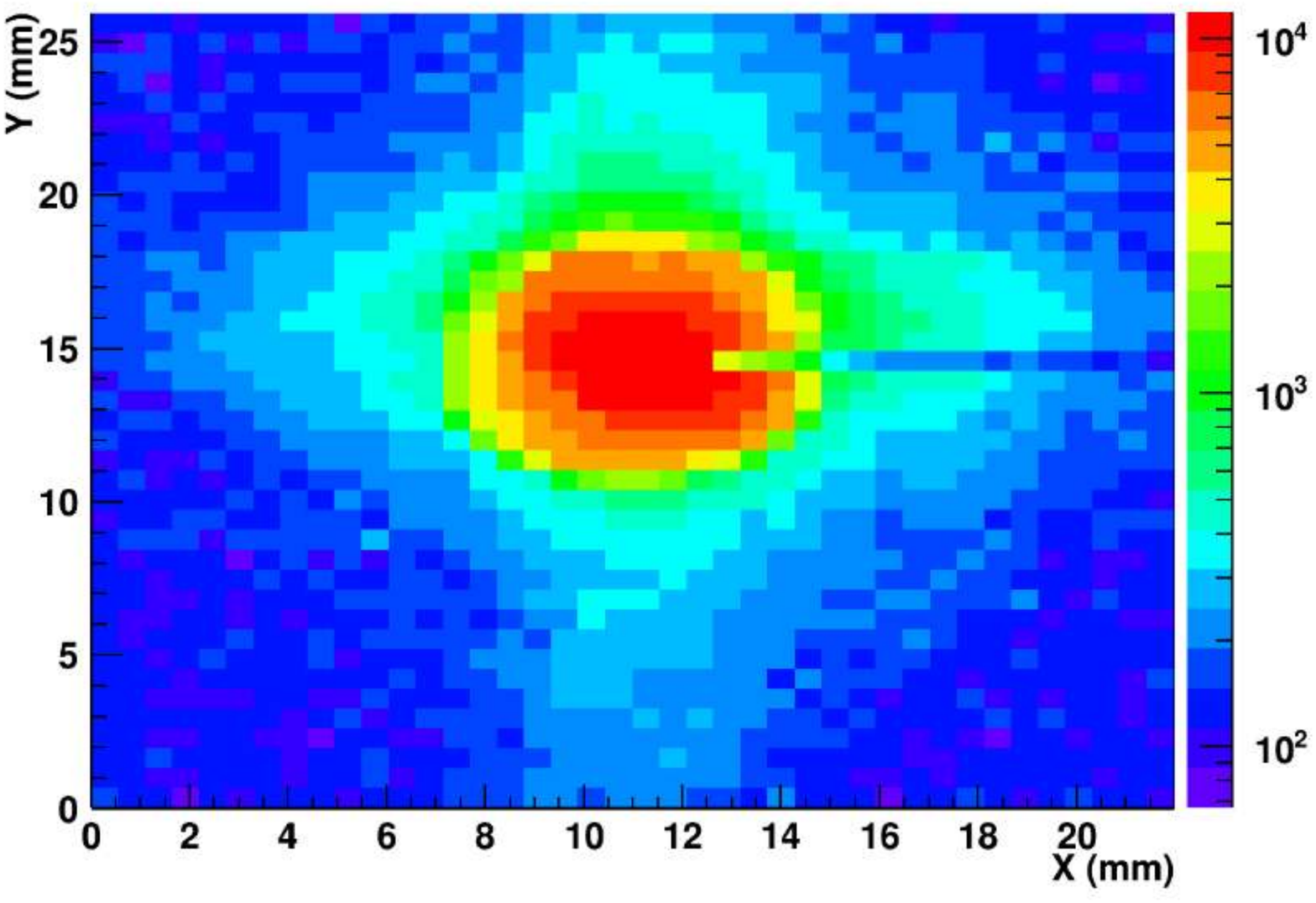}
  \end{minipage}%
  }%
  \subfigure[HPK SiPM in LAB]{
  \begin{minipage}{0.5\linewidth}
  \centering
   \includegraphics[width=8cm]{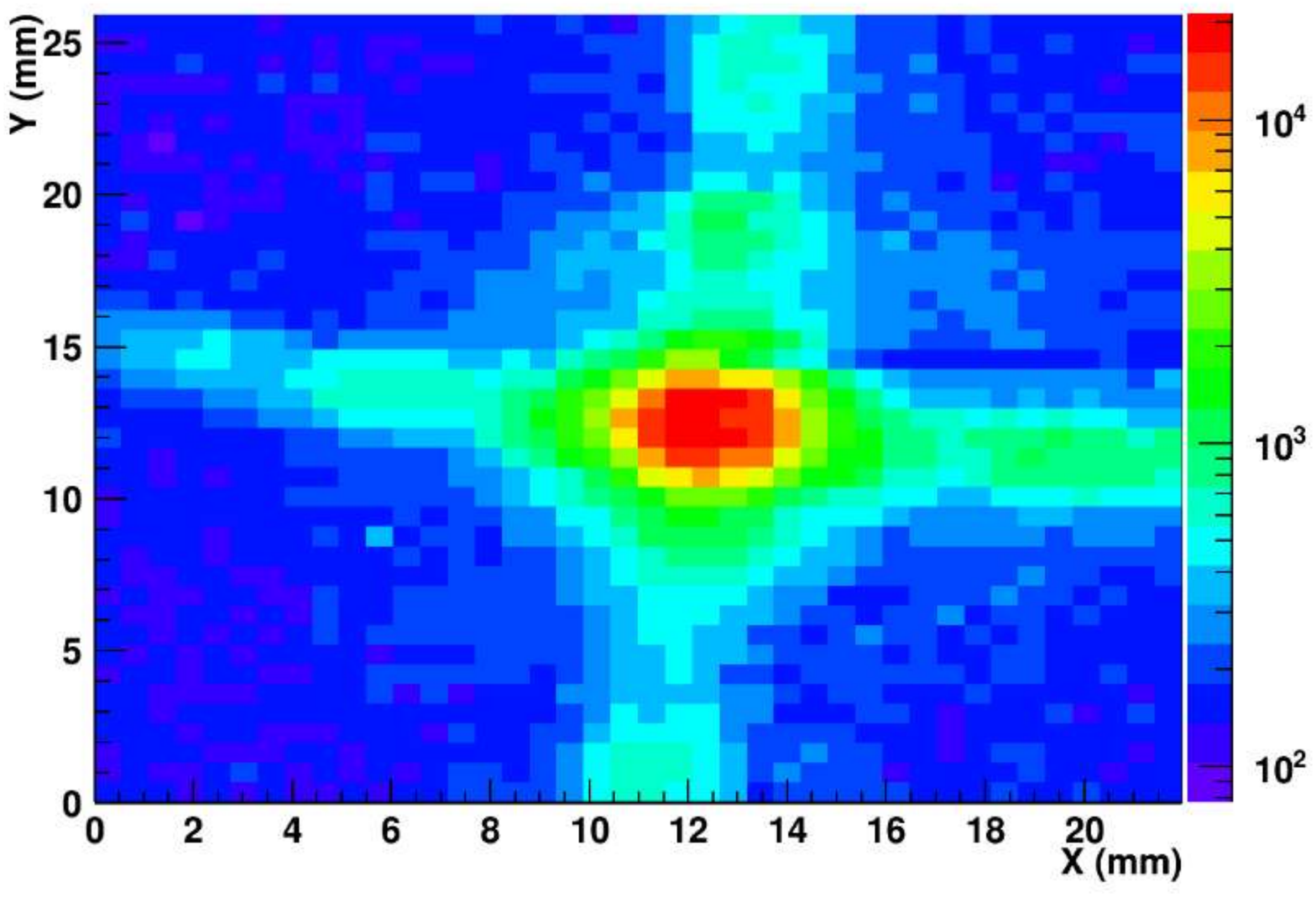}
  \end{minipage}%
  }%
  \caption{Profiles of reflected light of the FBK SiPM and the HPK SiPM in air and in LAB measured at the wavelength of 430 nm and the AOI of 50 degrees.}
  \label{diffuse}
\end{figure}

A fraction of diffuse reflectance is expected due to the microstructures on the SiPM surfaces, such as the guard rings, traces, and quenching resistors. Ideally, the profile of the reflected light can be measured by rotating the detector PD; however, an aperture must be installed in front of the PD to reduce the active area and obtain the fine structure of the profile. It is time consuming to complete a full scan for one profile. In this work, we place a CCD camera in the reflected light beams and measure their profiles for the two SiPMs. The active area of the CCD is 3.5 cm x 4 cm. The wavelength of incident light is selected as 430 nm. The profile is measured at the AOI of 50 degrees. The exposure time for each measurement is 1.2 seconds. Dark current data of the CCD are taken without light illumination on the CCD active area. Profiles of the reflected light in air and in LAB are displayed in Figure \ref{diffuse} for the two SiPMs without dark noise subtraction. The absolute values between different profiles cannot be directly compared because of the large light intensity fluctuation of the light beam, and it is difficult to correct these data using the reference PD within such a short exposure time. The broad distributions of reflected light in air and in LAB, shown in Figure \ref{diffuse}, indicate a fraction of a diffuse component in the SiPM reflections. The reflected light distributes within a broader region for the HPK SiPM compared with that of the FBK SiPM. When the SiPMs are immersed in LAB, the reflected light seems to be focused, and this can be explained by the good matching of the refractive indices between the acrylic vessel and LAB, which makes the vessel and LAB act like a lens. Cross shapes are clearly observed in the profiles of the two SiPMs, particularly when they are immersed in LAB. This feature is correlated to the microstructures on the SiPM surfaces. The shallow bands in plots are caused by the defect pixels of the CCD camera.

\section{Conclusion}
The reflectance is an important aspect in the evaluation of SiPMs, which can lead us to a better understanding of the large-scale SiPM-based photodetection systems and performance of SiPM devices. In the past, minimal knowledge was available regarding the reflectance of SiPMs in air or in any other media at visible wavelengths. In this work, a dedicated setup is designed and produced to measure the reflectance in air or in a liquid for various samples. The angular and spectral responses of the reflectance in air and in LAB are reported for the FBK SiPM (NUV-HD-lowCT) and the HPK SiPM (S14160-60-50HS), which are of interest for the TAO experiment. In general, there is a specular reflection ranging from 14\% to 24\% for the FBK SiPM in the region of measured wavelengths and AOIs, which is approximately 2 times that of the HPK SiPM. The reflectance in LAB is approximately 10\% lower than that measured in air. The effects of interference in the ARC are observed for both SiPMs. The profiles of reflected light beams are measured by a CCD camera and indicate that a fraction of diffuse reflections exists in addition to the specular component. Quantitative measurements will be conducted to study the diffuse reflections of SiPMs in the near future. 


\end{document}